\documentclass[12pt,a4paper]{article}

\makeatletter

\providecommand{\LyX}{L\kern-.1667em\lower.25em\hbox{Y}\kern-.125emX\@}
\newcommand{\noun}[1]{\textsc{#1}}

\makeatletter

\usepackage{latexsym}
\usepackage{amsmath}
\usepackage{amssymb}
\usepackage{amsthm}
\usepackage{amscd}

\usepackage{enumerate}

\swapnumbers
\theoremstyle{definition}
\newtheorem{thm}{Theorem}[section]
\newtheorem{defini}[thm]{Definition}
\newtheorem{prop}[thm]{Proposition}
\newtheorem{cor}[thm]{Corollary}

\newtheorem{lm}[thm]{Lemma}
\newtheorem{remark}[thm]{Remark}

\makeatletter
 
 \@addtoreset{equation}{section}
\makeatother

\def\rnum#1{\expandafter{\romannumeral #1}}
\def\Rnum#1{\uppercase\expandafter{\romannumeral #1}}

\makeatother

\begin{document}

\newpage\thispagestyle{empty}
{\topskip 2cm
\begin{center}
{\Huge\bf
On Quasifree Representations of Infinite Dimensional Symplectic Group}
\\
\bigskip\bigskip
\bigskip\bigskip
\bigskip\bigskip
{\Large Taku Matsui and Yoshihito Shimada}
\\
\bigskip
{\it Graduate School of Mathematics
\\
Kyushu University
\\
1-10-6 Hakozaki, Fukuoka 812-8581
\\
JAPAN}
\end{center}
\vfil
\noindent
We consider an infinite dimensional generalization
of Metaplectic representations (Weil representations)
for the (double covering of) symplectic group.
Given quasifree states of an infinite dimensional 
CCR algebra, projective unitary representations of 
the infinite dimensional symplectic group are constructed via
unitary implementors of Bogoliubov automorphisms.  
Complete classification of these representations
up to quasi-equivalence is obtained.
\noindent
\bigskip
\hrule
 \\
\noindent
{\small\tt e-mail: matsui@math.kyushu-u.ac.jp,  
shimada@math.kyushu-u.ac.jp}
\vfil}\newpage

\section{Introduction}
In this paper,  we consider unitary representations of 
an infinite symplectic group. The group $Sp(\infty)$ we deal with here is
the set of invertible operators $g=1+A$ where
$g$ preserves a symplectic form on an infinite dimensional vector space
and $A$ is of finite rank. This group is essentially same as
the inductive limit of the classical symplectic group
in the sense that the latter group is dense in operator
norm topology.
\par
First, we present a class of unitary representations of 
the Lie algebra  $sp(\infty)$ on GNS spaces of quasifree states of
the CCR (canonical commutation relations) algebra
(= the infinite dimensional Heisenberg algebra).
The construction of the representation is done 
in the same fashion as the metaplectic representation
(Weil representation) of finite dimensional groups.
(c.f. \cite{Lion-Vergne})
The representation constructed here is refered to as quasifree
represetation.
As in the finite dimensional case, our infinitesimal representations 
give rise to unitary representations  of the double covering of 
$Sp(\infty)$. 
The infinite dimensional CCR algebra has infinitely many mutually
non-equivalent representations, and  we obtain a huge number
of metaplectic representations of  $Sp(\infty)$.
This class of representations contains uncountably many irreducible
representations as well as non type I factor representations.
\bigskip
\noindent

The theory of unitary representation for infinite dimensional groups 
is a field of interplay between Ergodic Theory,  the measure theory of
infinite dimensional space, operators algebras, and mathematical
physics, in particular,  quantum field theory.
So far two classes of infinite dimensional groups are considered.
(1)  groups whose matrix elements are functions:
Examples are loop groups and the diffeomorphism group of the circle
(See \cite{Segal},  \cite{Carey} and the references therein),  and 
their higher dimensional analogue. 
 (2)  inductive limit of classical groups, $O(\infty)$ or $U(\infty)$.
See \cite{Borodin-Olshanski},  \cite{Pickrell} and 
\cite{Stratila-Voiculescu}.

So far the construction of unitary representations has been carried out 
in two ways.
 One way is to construct measures on an infinite dimensional space 
quasi-invariant under the group in question (\cite{Olshanski_Vershik}).
Another method is to use Fock spaces of the quantum field theory
and unitary implementors of Bogoliubov automorphisms.
(c.f. \cite{Carey}  and \cite{Stratila-Voiculescu}).

Turning into inductive limit $O(\infty)$ and $U(\infty)$ certain representations of these groups are closely connected with
the gauge invariant part of CAR (canonical anticommutation relations) 
algebra. In general the inductive limit procedure of compact groups
yields an inductive limit
of group $C^*$-algebras which are approximately finite dimensional.
Then there is a one to one correspondence of primitive ideals of the
AF algebra and  factor representations of the group.
The factor representation of $U(\infty)$
constructed on GNS spaces of quasifree states
of the CAR algebra corresponds to the U(1) gauge invariant 
part of the CAR algebra as the quotient by the primitive ideal.
(c.f. \cite{Stratila-Voiculescu}) In the same manner, spin representations
of $O(\infty)$ corresponds to the $Z_2$ invariant of  the CAR algebra.
Quasi-equivalence of quasifree states for gauge invariant CAR algebras
was investigated in \cite{Stratila-Voiculescu}, \cite{Matsui1} and 
\cite{Matsui} and these results leads to classification 
of representations of  $O(\infty)$ and $U(\infty)$ on GNS spaces
of quasifree states on CAR algebras.

In the same spirit, we can introduce quasifree representation 
(= Metaplectic or Weil representation) for $Sp(\infty)$  
with the aid of quasifree states of 
the CCR algebra. However, there is a crucial difference. 
The symplectic group $Sp(N)$ is non compact on one hand,
and the CCR algebra is an unbounded operator algebra.
The unitary representative of $O(\infty)$ or $U(\infty)$ is
an element of the gauge invariant CAR algebra while this is 
not the case for  $Sp(\infty)$.
In this sense, it is not correct that classification 
Metaplectic representations reduces to the representation 
theory of the $Z_2$ gauge invariant part of the CCR algebra. 
Nevertheless  we succeeded complete classification of 
generalized Metaplectic representations on GNS spaces 
associated with quasifree states of  the CCR algebra. 

The main result of this paper is Theorem 7.1 where we obtained
the complete  classification of quasifree representation
constructed in the GNS representations associated with quasifree states
of the CCR algebra.
To achieve our object we found it necessary to use Modular theory
of von Neumann algebra for quasifree states of CCR
algebras.  This machinery was established by H.Araki in
\cite{Araki-CCR-1}, \cite{Araki-CCR-2}, \cite{Araki-Fock-rep-CAR}
and \cite{Radon-Nikodym-theorem}.

Next we mention the organization of this paper. 
In Section 2 and 3 we introduce quasifree states of CCR algebras
 and Fock spaces in an abstract way.

The infinitesimal quasifree  representation of the Lie algebra
$sp(\infty)$ is defined in Section4.

 If the quasifree state of the CCR algebra
is pure, the associated representation of $sp(\infty)$ decomposes
into two mutually non-equivalent irreducible representations. 
This fact is proved in Section 5.

Section 6 is devoted to an analysis of von Neumann algebras
generated by  $sp(\infty)$. 
Using results of Section 6 our main result Theorem 7.1 is proved
in Section 7.

In the final section we show that our irreducible quasifree 
representation is extendible to a projective unitary representation 
of a larger symplectic group $Sp(P,\infty)$ where
$Sp(P,\infty)$ is a symplectic transformation commuting with
a fixed projection $P$ modulo Hilbert Schimidt class operators.
This result is closely connected with another result of D.Pickrell
in \cite{Pickrell} where he introduced the notion of
spherical representations and examined the same extension property. 

\section{Quasifree Representations of CCR algebra}

We briefly sketch GNS representations of CCR algebra associated to quasifree
states. 

\begin{defini} \textit{Let} \emph{\( K \)} \textit{be a complex vector space
and \( \gamma (f,g) \) be a non-degenerate hermitian form for \( f,g\in K \).
Let \( \Gamma  \) be an antilinear involution satisfying \( \Gamma ^{2}=1,\gamma (\Gamma f,\Gamma g)=-\gamma (g,f) \).
A self-dual CCR algebra} \textit{\emph{\( \mathfrak {A}(K,\gamma , \Gamma ) \)}}
\textit{is a complex} \textit{\emph{{*}-}}\textit{algebra generated by identity
\( 1 \) and \( \{B(f)\, |\, f\in K\} \) where \( B(f) \) is complex linear
in \( f\in K \) and satisfies \( B(f)^{*}=B(\Gamma f) \), \( B(f)^{*}B(g)-B(g)B(f)^{*}=\gamma (f,g) \).}\end{defini}

\begin{defini} \textit{A state \( \varphi  \) on} \textit{\emph{\( \mathfrak {A}(K,\gamma ,\Gamma ) \)}}
\textit{is a called quasifree state} \textit{\emph{}}\textit{if  
\begin{gather*}
 \varphi (B(f_{1})\ldots B(f_{2n-1}))=0,\\
 \varphi (B(f_{1})\ldots B(f_{2n}))=\sum _{\sigma \in \mathfrak {S}}\prod ^{n}_{j=1}\varphi (B(f_{\sigma (j)})B(f_{\sigma (n+j)})). 
\end{gather*}
\( \mathfrak {S} \) is the set of all permutations of \( \{1,2,\ldots ,n\} \)
satisfying} 
\[
\sigma (1)<\sigma (2)<\ldots <\sigma (n),\, \sigma (j)<\sigma (j+n),j=1,2,\ldots ,n.\]
 The cardinal number of \( \mathfrak {S} \) equals \( (2n)!2^{-n}(n!)^{-1} \).
\end{defini}

For any quasifree state \( \varphi  \), let
\[
S(f,g):=\varphi (B(f)^{*}B(g)).\]
Then a positive semi-definite hermitian form \( S(\cdot ,\cdot ) \) satisfies
\begin{equation}
\label{eq:S(f,g)}
S(f,g)-S(\Gamma g,\Gamma f)=\gamma (f,g).
\end{equation}
 Conversely, a positive semi-definite hermitian form \( S:K\times K\rightarrow \mathbf{C} \)
satisfying \eqref{eq:S(f,g)} is given. Then there exists an unique quasifree
state \( \varphi _{S} \) on \emph{\( \mathfrak {A}(K,\gamma ,\Gamma ) \)}
such that
\[
\varphi _{S}(B(f)^{*}B(g))=S(f,g)\]
for all \( f,g\in K \). That is to say, a quasifree state is completely specified
by a positive semi-definite hermitian form satisfying \eqref{eq:S(f,g)}. (See
\cite{Araki-CCR-1}.) 

We define a bounded operator ``\( S \)'' induced by a positive semi-definite
hermitian form \( S \) satisfying \eqref{eq:S(f,g)}. Due to the non-degeneracy
of \( \gamma  \), a hermitian form
\[
(f,g)_{S}:=S(f,g)+S(\Gamma g,\Gamma f)\]
 is positive definite. In other words \( (\cdot ,\cdot )_{S} \) is an inner
product. Let \( K_{S} \) be the completion of \( K \) with respect to \( (\cdot ,\cdot )_{S} \).
Then there exists a bounded operator \( S \) on \( K_{S} \) such that
\[
S(f,g)=(f,Sg)_{S}\]
 for all \( f,g\in K \). Let \( \Gamma _{S} \) be an antiunitary involution
on \( K_{S} \) such that \( \Gamma _{S}f=\Gamma f \) for all \( f\in K \).
The bounded operator \( S \) satisfies \( S^{*}=S \), \( \Gamma _{S}S\Gamma _{S}=1-S \)
and \( 0\leq S\leq 1 \). \( \gamma _{S}:=2S-1 \) satisfies \( \gamma (f,g)=(f,\gamma _{S}g)_{S} \)
for all \( f,g\in K \). 

If the bounded operator \( S \) on \( K_{S} \) induced by a positive semi-definite
hermitian form satisfying \eqref{eq:S(f,g)} is a projection, we call \( S \)
a \emph{basis projection}. 

Let \( (\mathcal{H}_{S},\pi _{S},\Omega _{S}) \) be a GNS representation of
CCR algebra \( \mathfrak {A}(K,\gamma ,\Gamma ) \) associated to \( \varphi _{S} \).
The Hilbert space given by GNS construction is abstract, however in case that
\( S \) is a basis projection, it can be written concretely.

Let \( L \) be a Hilbert space and consider the Boson Fock space :
\begin{gather*}
 \mathcal{F}_{\mathrm{b}}(L):=\bigoplus ^{\infty }_{n=0}\otimes ^{n}_{\mathrm{s}}L,\quad \otimes ^{0}_{\mathrm{s}}L:=\mathbf{C},\quad \Psi =1,\\
 \left\langle f_{1}\otimes _{\mathrm{s}}\ldots \otimes _{\mathrm{s}}f_{n},g_{1}\otimes _{\mathrm{s}}\ldots \otimes _{\mathrm{s}}g_{m}\right\rangle =\delta _{mn}\frac{1}{n!}\sum _{\sigma \in \mathfrak {S}_{n}}\prod _{j=1}^{n}(f_{\sigma (j)},g_{\sigma (j)}).
\end{gather*}
\( \otimes _{\mathrm{s}} \) is the symmetric tensor product and \( \mathfrak {S}_{n} \)
is the set of all permutations of \( \{1,2,\ldots ,n\} \). Now we define annihilation
operators \( b(f),f\in L \) on \( \mathcal{F}_{\mathrm{b}}(L) \) as follows
:
\begin{gather*}
 b(f)f_{1}\otimes _{\mathrm{s}}\ldots \otimes _{\mathrm{s}}f_{n}:=\frac{1}{\sqrt{n}}\sum _{j=1}^{n}(f,f_{j})f_{1}\otimes _{\mathrm{s}}\ldots \otimes _{\mathrm{s}}f_{j-1}\otimes _{\mathrm{s}}f_{j+1}\ldots \otimes _{\mathrm{s}}f_{n},\\
 b(f)\Psi :=0 
\end{gather*}
and creation operators \( b^{\dagger }(f),f\in L \) on \( \mathcal{F}_{\mathrm{b}}(L) \)
as follows : \emph{
\begin{gather*}
 b^{\dagger }(f_{1})f_{2}\otimes _{\mathrm{s}}\ldots \otimes _{\mathrm{s}}f_{n+1}:=\sqrt{n+1}f_{1}\otimes _{\mathrm{s}}\ldots \otimes _{\mathrm{s}}f_{n+1},\\
 b^{\dagger }(f)\Psi :=f. 
\end{gather*}
} \begin{lm}\label{thm:CCR-Fock-rep} \emph{Suppose that} \( S:K_{S}\rightarrow K_{S} \)
\emph{is a basis projection}.

\begin{enumerate}
\item \emph{\( b^{\dagger }(f),b(g),f,g\in SK_{S} \) are closable operators}. \emph{Let
\( \overline{A} \) be the closure of operator \( A \)}. \emph{The finite particle
vector subspace of \( \mathcal{F}_{\mathrm{b}}(SK_{S}) \) is a core for all
\( \overline{b^{\dagger }(f)},\overline{b(g)},f,g\in SK_{S} \)}. 
\item \emph{Due to} (1), \emph{we can define the addition and multiplication of creation
and annihilation operators on the finite particle vector subspace of \( \mathcal{F}_{\mathrm{b}}(SK_{S}) \).
Let \( \mathfrak {A}_{\mathbf{CCR}}(SK_{S}) \) be a} {*}-\emph{algebra} \emph{generated
by all annihilation and creation operators}. \emph{Let} \emph{\( \alpha (S):\mathfrak {A}(K,\gamma ,\Gamma )\rightarrow \mathfrak {A}_{\mathbf{CCR}}(SK_{S}) \)
be a} {*}-\emph{homomorphism} \emph{satisfying the following relation} : \emph{
\[
\alpha (S)(B(f)):=b^{\dagger }(Sf)+b(S\Gamma f),f\in K.\]
 Then \( (\mathcal{F}_{\mathrm{b}}(SK_{S}),\alpha (S),\Psi ) \) is a} {*}-\emph{representation
of CCR algebra \( \mathfrak {A}(K,\gamma ,\Gamma ) \). Moreover}, \emph{it
is unitary equivalent to the GNS representation} \( (\mathcal{H}_{S},\pi _{S},\Omega _{S}). \) 
\end{enumerate}
\end{lm} \begin{proof} (1) See chapter X section 7 of \cite{Reed-Simon-2}.
(2) An unitary operator \( u:\mathcal{H}_{S}\rightarrow \mathcal{F}_{\mathrm{b}}(SK_{S}), \)
\( \pi _{S}(X)\Omega _{S}\mapsto \alpha (S)(X)\Psi  \) satisfies \( u^{*}\alpha (S)(X)u=\pi _{S}(X) \)
for all \( X\in \mathfrak {A}(K,\gamma ,\Gamma ) \).\end{proof}

By Lemma \ref{thm:CCR-Fock-rep}, we call \( \pi _{S} \) a \emph{Fock representation}
and \( \varphi _{S} \) a \emph{Fock state} if \( S \) is a basis projection.
In case that \( \varphi _{S} \) is a Fock state, we have the following important
lemma. (See Lemma 5.4 and 5.5 of \cite{Araki-CCR-1}.)

\begin{lm}\label{thm:Fock-rep-property} \emph{Suppose that} \( S:K_{S}\rightarrow K_{S} \)
\emph{is a basis projection}. \emph{}

\begin{enumerate}
\item \( \pi _{S}(B(f)),f\in \mathrm{Re}K \) \emph{is an essentially self-adjoint
operator and set}
\[
W_{S}(f):=\exp \left( i\overline{\pi _{S}(B(f)}\right) .\]
 \emph{Then \( W_{S}(f) \) satisfies the following relations} : 
\[
W_{S}(f_{1})W_{S}(f_{2})=\exp \left( -\frac{1}{2}\gamma (f_{1},f_{2})\right) W_{S}(f_{1}+f_{2}).\]
 
\item \emph{If \( f\in \mathrm{Re}K_{S} \)}, \emph{we define \( W_{S}(f) \) via
the following limit
\begin{equation}
\label{eq:limW(f_n)}
W_{S}(f):=\mathrm{s}\frac{\, \, }{\, \, }\lim _{n\rightarrow \infty }W_{S}(f_{n})
\end{equation}
 where} \( \{f_{n}\} \) \emph{is a sequence in} \( \mathrm{Re}K \) \emph{satisfying}
\( \left\Vert f-f_{n}\right\Vert \rightarrow 0 \). \emph{Note that the limit}
\eqref{eq:limW(f_n)} \emph{does not depend on the choice of} \( \{f_{n}\} \).
\item \emph{Let} \( \mathrm{Re}K_{S}:=\{f\in K_{S}\, |\, \Gamma _{S}f=f\} \)\emph{.
The restriction of} \( (\cdot ,\cdot )_{S} \) \emph{to \( \mathrm{Re}K_{S} \)
is an inner product of \( \mathrm{Re}K_{S} \).} \( f\mapsto W_{S}(f) \) \emph{is
continuous with respect to the norm on} \( \mathrm{Re}K_{S} \) \emph{and the
strong operator topology of bounded operators on} \( \mathcal{H}_{S} \).
\item \emph{Let \( L \) be a subspace of \( \mathrm{Re}K_{S} \). Let \( L^{\vee } \)
be the set of vectors \( f\in \mathrm{Re}K_{S} \) such that \( (f,\gamma _{S}g)_{S}=0 \)
for all \( g\in L \) and let \( \overline{L} \) be the closure of \( L \)
in} \( \mathrm{Re}K_{S} \). \emph{Let \( \mathcal{R}_{S}(L) \) be a von Neumann
algebra generated by} \( W_{S}(f),f\in L. \) \emph{Then we obtain the following
relations} : 

\begin{enumerate}
\item \( \mathcal{R}_{S}(L)=\mathcal{R}(\overline{L}), \)
\item \( \mathcal{R}_{S}(L)'=\mathcal{R}_{S}(L^{\vee }), \)
\item \( \{\mathcal{R}_{S}(L_{1})\cup \mathcal{R}_{S}(L_{2})\}''=\mathcal{R}_{S}(L_{1}+L_{2}), \)
\item \( \mathcal{R}_{S}(L_{1})\cap \mathcal{R}_{S}(L_{2})=\mathcal{R}_{S}(\overline{L_{1}}\cap \overline{L_{2}}). \)
\end{enumerate}
\end{enumerate}
\end{lm}

We introduce an another hermitian form \( \widehat{\gamma }_{S} \) on \( K_{S}\oplus K_{S} \)
via the following relation :
\[
\widehat{\gamma }_{S}(f_{1}\oplus g_{1},f_{2}\oplus g_{2}):=(f_{1},\gamma _{S}f_{2})_{S}-(g_{1},\gamma _{S}g_{2})_{S}\]
 for all \( f_{i},g_{i}\in K_{S} \). Set \( \widehat{\Gamma }_{S}:=\Gamma _{S}\oplus \Gamma _{S} \).
Then \( \widehat{\gamma }_{S} \) satisfies \( \hat{\gamma }_{S}(\hat{\Gamma }_{S}h_{1},\hat{\Gamma }_{S}h_{2})=-\hat{\gamma }_{S}(h_{1},h_{2}) \)
for all \( h_{i}\in K_{S}\oplus K_{S} \). 
\[
P_{S}(f_{1}\oplus f_{2},g_{1}\oplus g_{2}):=(f_{1},Sf_{2}+\sqrt{S(1-S)}g_{2})_{S}+(g_{1},\sqrt{S(1-S)}f_{2}+(1-S)g_{2})_{S}\]
is a positive semi-definite hermitian form on \( K_{S}\oplus K_{S} \) satisfying
\[
P_{S}(h_{1},h_{2})-P_{S}(\widehat{\Gamma }_{S}h_{2},\hat{\Gamma }_{S}h_{1})=\hat{\gamma }_{S}(h_{1},h_{2}).\]
We denote the completion of \( K_{S}\oplus K_{S} \) with respect to the inner
product \( (h_{1},h_{2})_{P_{S}}:=P_{S}(h_{1},h_{2})+P_{S}(\hat{\Gamma }_{S}h_{2},\hat{\Gamma }_{S}h_{1}) \)
by \( K_{P_{S}} \). 

\begin{lm}\label{thm:P_S-is-a-projection} \emph{The bounded operator} \emph{\( P_{S} \)
on} \( K_{P_{S}} \) \emph{satisfying} \( P_{S}(h_{1},h_{2})=(h_{1},P_{S}h_{2})_{P_{S}} \)
\emph{for all} \( h_{i}\in K_{S}\oplus K_{S} \) \emph{is a basis projection}.
\end{lm}
\begin{proof} Let \( D(\gamma ^{-1}_{S}) \) be a domain of \( \gamma ^{-1}_{S} \).
By the non-degeneracy of \( \gamma  \), \( D(\gamma ^{-1}_{S}) \) is a dense
set of \( K_{S} \). Let \( \widehat{\gamma }_{S}(f_{1}\oplus g_{1},f_{2}\oplus g_{2})=(f_{1}\oplus g_{1},h\oplus k)_{P_{S}} \).
Then \( \gamma _{S}f_{2}=h+2\sqrt{S(1-S)}k \), \( -\gamma _{S}g_{2}=2\sqrt{S(1-S)}h+k \).
If \( f_{2},g_{2}\in D(\gamma ^{-1}_{S}) \), we have
\[
\gamma _{P_{S}}(f_{2}\oplus g_{2})=h\oplus k=\gamma _{S}^{-1}(f_{2}+2\sqrt{S(1-S)}g_{2})\oplus -\gamma _{S}^{-1}(2\sqrt{S(1-S)}f_{2}+g_{2}).\]
By \( P_{S}=\frac{1}{2}(\gamma _{P_{S}}+1) \), \( P_{S} \) is written explicitly
on \( D(\gamma ^{-1}_{S})\oplus D(\gamma ^{-1}_{S}) \) as follows : 
\[
P_{S}(f\oplus g)=\gamma ^{-1}_{S}(Sf+\sqrt{S(1-S)}g)\oplus -\gamma ^{-1}_{S}(\sqrt{S(1-S)}f+(1-S)g).\]
It is easily checked that \( P_{S} \) is a projection on \( K_{P_{S}} \).\end{proof} 

\emph{Remark}. Let \( L \) be a dense set of \( K_{S} \) with respect to \( (\cdot ,\cdot )_{S} \),
then \( L\oplus L \) is a dense set of \( K_{S}\oplus K_{S} \) with respect
to \( (\cdot ,\cdot )_{P_{S}} \). Indeed, for any \( f\oplus g\in K_{S}\oplus K_{S} \),
there exist \( f_{n},g_{n}\in L \) such that \( \parallel f_{n}-f\parallel _{S},\parallel g_{n}-g\parallel _{S}\rightarrow 0(n\rightarrow \infty ) \).
By the following equation 
\begin{equation}
\label{eq:relation_between_<,>(S)_and_<,>(P_S)}
\parallel f\oplus g\parallel ^{2}_{P_{S}}=\parallel \sqrt{S}f+\sqrt{1-S}g\parallel ^{2}_{S}+\parallel \sqrt{1-S}f+\sqrt{S}g\parallel ^{2}_{S},
\end{equation}
we have \( \parallel (f_{n}\oplus g_{n})-(f\oplus g)\parallel _{P_{S}}\rightarrow 0(n\rightarrow \infty ) \).

\medskip

By Lemma \ref{thm:P_S-is-a-projection}, \( \varphi _{P_{S}} \) is a Fock state
on CCR algebra \( \mathfrak {A}(K_{S}\oplus K_{S},\widehat{\gamma }_{S},\widehat{\Gamma }_{S}). \)
We denote a GNS representation of CCR algebra \( \mathfrak {A}(K_{S}\oplus K_{S},\hat{\gamma }_{S},\hat{\Gamma }_{S}) \)
associated to \( \varphi _{P_{S}} \) by \( (\mathcal{H}_{P_{S}},\pi _{P_{S}},\Omega _{P_{S}}) \)
. The following corollary is a consequence of the direct application of lemma
\ref{thm:Fock-rep-property} to Fock representation \( (\mathcal{H}_{P_{S}},\pi _{P_{S}},\Omega _{P_{S}}) \). 

\begin{cor}\label{thm:factoriarity_of_R(ReK+0)} \( \mathcal{R}_{P_{S}}(\mathrm{Re}K_{S}\oplus 0)'=\mathcal{R}_{P_{S}}(0\oplus \mathrm{Re}K_{S}) \) \emph{and}
\( \mathcal{R}_{P_{S}}(\mathrm{Re}K_{S}\oplus 0) \) \emph{is a factor}.\end{cor}

\begin{remark}\label{thm:Remark}Let \( \alpha :\mathfrak {A}(K,\gamma ,\Gamma )\rightarrow \mathfrak {A}(K_{S}\oplus K_{S},\widehat{\gamma }_{S},\widehat{\Gamma }_{S}) \)
be a {*}-homomorphism defined by \( \alpha (B(f))=B(f\oplus 0) \) and \( u_{\alpha }:\mathcal{H}_{S}\rightarrow \mathcal{H}_{P_{S}} \)
be a linear operator defined by \( u_{\alpha }(\pi _{S}(A)\Omega _{S})=\pi _{P_{S}}(\alpha (A))\Omega _{P_{S}} \)
for all \( A\in \mathfrak {A}(K,\gamma ,\Gamma ). \) Then \( u_{\alpha } \)
preserves the inner product. In fact, since \( \varphi _{P_{S}} \) and \( \varphi _{S} \)
are quasifree states and
\[
\varphi _{P_{S}}(B(f\oplus 0)^{*}B(g\oplus 0))=P_{S}(f\oplus 0,g\oplus 0)=S(f,g)=\varphi _{S}(B(f)^{*}B(g))\]
for all \( f,g\in K \), we have \( \varphi _{P_{S}}(\alpha (A))=\varphi _{S}(A) \)
for all \( A\in \mathfrak {A}(K,\gamma ,\Gamma ). \) If \( X\) and \(
Y\) are elements of \( \mathfrak {A}(K,\gamma ,\Gamma ) \) and set \(
A:=X^{*}Y \), then 
\[
\left\langle u_{\alpha }\pi _{S}(X)\Omega _{S},u_{\alpha }\pi _{S}(Y)\Omega _{S}\right\rangle =\varphi _{P_{S}}(\alpha (X^{*}Y))=\varphi _{S}(X^{*}Y)=\left\langle \pi _{S}(X)\Omega _{S},\pi _{S}(Y)\Omega _{S}\right\rangle .\]
If we identify \( u_{\alpha }\mathcal{H}_{S} \) with \( \mathcal{H}_{S} \),
\( \mathcal{H}_{S} \) is a closed subspace of \( \mathcal{H}_{P_{S}} \) and
\( \mathcal{H}_{S}=\mathcal{F}_{\mathrm{b}}(P_{S}(K_{S}\oplus 0)) \). Moreover,
since \( u_{\alpha }\pi _{S}(A)=\pi _{P_{S}}(\alpha (A))u_{\alpha } \) on \( D(\pi _{S}):=\pi _{S}(\mathfrak {A}(K,\gamma ,\Gamma ))\Omega _{S} \),
we can identify \( \pi _{S}(A) \) with \( \pi _{P_{S}}(\alpha (A))|D(\pi _{S}) \)
for all \( A\in \mathfrak {A}(K,\gamma ,\Gamma ). \)\end{remark}

\section{Fock Space and Exponential Vectors}

In Remark \ref{thm:Remark}, \( \mathcal{H}_{S} \) is regarded as a closed
subspace of \( \mathcal{H}_{P_{S}} \). We explain the point in detail. 

Let \( L_{1},L_{2} \) be Hilbert spaces and \( L:=L_{1}\oplus L_{2} \) and
\( e(u):=\sum _{n=0}^{\infty }(\sqrt{n!})^{-1}\otimes _{\mathrm{s}}^{n}u \)
for all \( u\in L \). We call \( e(u) \) an \emph{exponential vector}. 

\begin{lm}\label{thm:decomposition_of_Fock_space} If \( u_{1}\in L_{1} \)
and \( u_{2}\in L_{2} \), then there exists an unique unitary operator \( U:\mathcal{F}_{\mathrm{b}}(L)\rightarrow \mathcal{F}_{\mathrm{b}}(L_{1})\otimes \mathcal{F}_{\mathrm{b}}(L_{2}) \)
such that \( Ue(u_{1}+u_{2})=e(u_{1})\otimes e(u_{2}) \). This shows
\[
\mathcal{F}_{\mathrm{b}}(L)=\mathcal{F}_{\mathrm{b}}(L_{1})\otimes \mathcal{F}_{\mathrm{b}}(L_{2}).\]
(See chapter II section 19 of \cite{K.R.Parthasarathy}.)\end{lm}

\begin{lm}\label{thm:decomposition_of_PK}

\( P_{S}K_{P_{S}}=[P_{S}(K_{S}\oplus 0)]\oplus [0\oplus E_{S}(\{0\})K_{S}]=[P_{S}(0\oplus K_{S})]\oplus [E_{S}(\{1\})K_{S}\oplus 0] \)
\emph{where} \( E_{S}(B) \) \emph{is the spectral projection of \( S \) for
a Borel set} \( B\subset \mathbf{R} \). \end{lm}
\begin{proof} The restriction of \( (\cdot ,\cdot )_{P_{S}} \) to \( P_{S}K_{P_{S}} \)
is an inner product of \( P_{S}K_{P_{S}} \). Let \( [P_{S}(K_{S}\oplus 0)]^{\perp } \)
be the orthogonal complement of \( P_{S}(K_{S}\oplus 0) \) in \( P_{S}K_{S} \).
Let \( u\oplus v\in [P_{S}(K_{S}\oplus 0)]^{\perp } \), then we have 
\[
0=(u\oplus v,P_{S}(f\oplus 0))_{P_{S}}=(Su+\sqrt{S(1-S)}v,f)_{S}\]
 for all \( f\in K_{S} \). This implies \( \sqrt{S}(\sqrt{S}u+\sqrt{1-S}v)=0 \),
i.e. \( Su+\sqrt{1-S}v\in E_{S}(\{0\})K_{S} \). By \( P_{S}(u\oplus v)=u\oplus v \)
and \( \gamma ^{-1}_{S}=-1,\sqrt{1-S}=1,\sqrt{S}=0 \) on \( E_{S}(\{0\})K_{S} \),
we have 
\[
u\oplus v=P(u\oplus v)=0\oplus (\sqrt{S}u+\sqrt{1-S}v).\]
Thus \( u=0 \), \( v\in E_{S}(\{0\})K_{S} \). We obtain \( [P_{S}(K_{S}\oplus 0)]^{\perp }\subset 0\oplus E_{S}(\{0\})K_{S}. \)
Converse relation is seen from direct computation. \end{proof}

From Lemma \ref{thm:decomposition_of_Fock_space}, Lemma \ref{thm:decomposition_of_PK}
and Remark \ref{thm:Remark}, we obtain the factorization of the Fock space
:
\begin{equation}
\label{eq:H(P_S)_equal_H(S)xL(S)}
\mathcal{H}_{P_{S}}=\mathcal{H}_{S}\otimes \mathcal{L}_{S},
\end{equation}
 \( \mathcal{L}_{S}:=\mathcal{F}_{\mathrm{b}}(0\oplus E_{S}(\{0\})K_{S}) \).
In particular, if \( 0<S<1 \), then \( \mathcal{H}_{P_{S}}=\mathcal{H}_{S}. \)

Let \( L \) be a Hilbert space and 
\[
\mathcal{F}_{\mathrm{b}}^{+}(L):=\bigoplus _{n=0}^{\infty }\otimes _{\mathrm{s}}^{2n}L,\quad \mathcal{F}_{\mathrm{b}}^{-}(L):=\bigoplus _{n=0}^{\infty }\otimes _{\mathrm{s}}^{2n+1}L.\]
We call \( \mathcal{F}_{\mathrm{b}}^{+}(L) \) the \emph{even part of the Boson
Fock space} \( \mathcal{F}_{\mathrm{b}}(L) \) and \( \mathcal{F}_{\mathrm{b}}^{-}(L) \)
the \emph{odd part of the Boson Fock space} \( \mathcal{F}_{\mathrm{b}}(L) \).

\begin{lm} \emph{Let
\[
e^{+}(u):=\bigoplus _{n=0}^{\infty }\frac{1}{\sqrt{(2n)!}}\otimes _{\mathrm{s}}^{2n}u,\quad e^{-}(u):=\bigoplus _{n=0}^{\infty }\frac{1}{\sqrt{(2n+1)!}}\otimes _{\mathrm{s}}^{2n+1}u\]
for all \( u\in L \)}. \emph{Let \( u_{j}\in L \), \( j=1,2,\ldots ,N \),
\( N\in \mathbf{N} \) satisfy \( u_{i}\neq \pm u_{j}(i\neq j) \). Then \( \{e^{\sigma }(u_{j})\}_{j=1}^{N} \)
is linearly independent. Moreover, \( \{e^{\sigma }(u)\, |\, u\in L\} \) generates
\( \mathcal{F}^{\sigma }_{\mathrm{b}}(L) \) where \( \sigma =+ \) or \( - \)}.
\end{lm}
\begin{proof} First, we prove the linear independence of \emph{\( \{e^{+}(u_{j})\}_{j=1}^{N} \)}.
Let \( \sum _{j=1}^{N}\alpha _{j}e^{+}(u_{j})=0,\, \alpha _{j}\in \mathbf{C} \).
We have \( 0=\left\langle e^{+}(x),\sum _{j=1}^{N}\alpha _{j}e^{+}(u_{j})\right\rangle =\sum _{j=1}^{N}\alpha _{j}\cosh (x,u_{j}) \)
for all \( x\in L \). If there exist \( i,j(i\neq j) \) such that \( (x,u_{i})^{2}=(x,u_{j})^{2} \)
for all \( x\in L \), then \( u_{i}=u_{j} \) or \( u_{i}=-u_{j} \). Thus
there exists \( x_{0}\in L \) such that \( (x_{0},u_{i})^{2}\neq (x_{0},u_{j})^{2} \)
for all \( i,j(i\neq j) \). Let \( x=\overline{z}x_{0},\, z\in \mathbf{C},\, \beta _{j}:=(x_{0},u_{j}) \),
then we have \( \sum _{j=1}^{N}\alpha _{j}\cosh (\beta _{j}z)=0 \) for all
\( z\in \mathbf{C} \). By the \( 2k \) th differential \( \sum _{j=1}^{N}\alpha _{j}\beta _{j}^{2k}\cosh (\beta _{j}z)=0 \),
\( k=1,2,\ldots ,N-1 \), we have 
\begin{equation}
\label{eq:van-der-Mondian}
\left( \begin{array}{cccc}
1 & 1 & \cdots  & 1\\
\beta ^{2}_{1} & \beta ^{2}_{2} & \cdots  & \beta ^{2}_{N}\\
\vdots  & \vdots  & \ddots  & \vdots \\
\beta ^{2(N-1)}_{1} & \beta ^{(N-1)}_{2} & \cdots  & \beta ^{2(N-1)}_{N}
\end{array}\right) \left( \begin{array}{c}
\alpha _{1}\cosh (\beta _{1}z)\\
\alpha _{2}\cosh (\beta _{2}z)\\
\vdots \\
\alpha _{N}\cosh (\beta _{N}z)
\end{array}\right) =\left( \begin{array}{c}
0\\
0\\
\vdots \\
0
\end{array}\right) 
\end{equation}
for all \( z\in \mathbf{C} \). Since the matrix of the left hand side of \eqref{eq:van-der-Mondian}
is a Vandermonde matrix and \( \beta _{i}^{2}\neq \beta _{j}^{2}(i\neq j) \),
its determinant does not vanish and we obtain \( \alpha _{j}\cosh (\beta _{j}z)=0 \)
for all \( z\in \mathbf{C} \) and \( j. \) Therefore, \( \alpha _{j}=0 \)
for all \( j \). Linear independence of \( \{e^{-}(u_{j})\}_{j=1}^{N} \) is
verified by putting ``\( \sinh  \)'' to the place of ``\( \cosh  \)''
in the above proof. 

We prove the second part of this lemma. The case of \( \{e^{+}(u)\, |\, u\in L\} \)
is the same as the proof of Proposition 19.4 of \cite{K.R.Parthasarathy}. Since
\( \mathcal{F}_{\mathrm{b}}(L) \) is generated by exponential vectors \( e(u),u\in L \)
and \( e(u)=e^{+}(u)\oplus e^{-}(u) \), \( \{e^{-}(u)\, |\, u\in L\} \) generates
\( \mathcal{F}^{-}_{\mathrm{b}}(L) \). \end{proof}
\begin{lm} \emph{Let \( L_{1},L_{2} \) be Hilbert spaces. Then there exist
unitary operators \( U_{+},U_{-} \) such that} 
\begin{gather}
 U_{+}:\mathcal{F}_{\mathrm{b}}^{+}(L_{1}\oplus L_{2})\rightarrow [\mathcal{F}_{\mathrm{b}}^{+}(L_{1})\otimes \mathcal{F}_{\mathrm{b}}^{+}(L_{2})]\oplus [\mathcal{F}_{\mathrm{b}}^{-}(L_{1})\otimes \mathcal{F}_{\mathrm{b}}^{-}(L_{2})],\label{eq:U(+)} \\
 U_{+}e^{+}(x_{1}\oplus x_{2}):=[e^{+}(x_{1})\otimes e^{+}(x_{2})]\oplus [e^{-}(x_{1})\otimes e^{-}(x_{2})] \label{eq:decomposition_of_Fock_space(+)} 
\end{gather}
\emph{and} 
\begin{gather}
 U_{-}:\mathcal{F}_{\mathrm{b}}^{-}(L_{1}\oplus L_{2})\rightarrow [\mathcal{F}_{\mathrm{b}}^{+}(L_{1})\otimes \mathcal{F}_{\mathrm{b}}^{-}(L_{2})]\oplus [\mathcal{F}_{\mathrm{b}}^{-}(L_{1})\otimes \mathcal{F}_{\mathrm{b}}^{+}(L_{2})], \label{eq:U(-)} \\
 U_{-}e^{+}(x_{1}\oplus x_{2}):=[e^{+}(x_{1})\otimes e^{-}(x_{2})]\oplus [e^{-}(x_{1})\otimes e^{+}(x_{2})]. \label{eq:decomposition_of_Fock_space(-)} 
\end{gather}
\end{lm}

\begin{proof} Let \( U \) be the unitary operator determined by Lemma \ref{thm:decomposition_of_Fock_space}.
Let \( U_{\sigma}:=U|\mathcal{F}_{\mathrm{b}}^{\sigma}(L_{1}\oplus
L_{2}) \) where \( \sigma =+ \) or \( - \). 
For all \( x_{1},y_{1}\in L_{1} \) and \( x_{2},y_{2}\in L_{2} \),
\[\begin{split}
\langle U_{+}e^{+}(x_{1}+x_{2}),U_{+}e^{+}(y_{1}+y_{2})\rangle 
 & = \left\langle Ue^{+}(x_{1}+x_{2}),Ue^{+}(y_{1}+y_{2})\right\rangle \\
 & = \left\langle e^{+}(x_{1}+x_{2}),e^{+}(y_{1}+y_{2})\right\rangle \\
 & = \cosh ((x_{1},y_{1})+(x_{2},y_{2}))\\
 & = \cosh (x_{1},y_{1})\cosh (x_{2},y_{2})+\sinh (x_{1},y_{1})\sinh (x_{2},y_{2})\\
 & = \left\langle e^{+}(x_{1}),e^{+}(y_{1})\right\rangle \left\langle e^{+}(x_{2}),e^{+}(y_{2})\right\rangle \\
 &  \qquad +\left\langle e^{-}(x_{1}),e^{-}(y_{1})\right\rangle \left\langle e^{-}(x_{2}),e^{-}(y_{2})\right\rangle \\
 & = \big \langle [e^{+}(x_{1})\otimes e^{+}(x_{2})]\oplus [e^{-}(x_{1})\otimes e^{-}(x_{2})],\\
 &  \qquad [e^{+}(y_{1})\otimes e^{+}(y_{2})]\oplus [e^{-}(y_{1})\otimes e^{-}(y_{2})]\big \rangle 
\end{split}\]
and
\[\begin{split}
\mathcal{F}_{\mathrm{b}}^{+}(L_{1}\oplus L_{2})\oplus
\mathcal{F}_{\mathrm{b}}^{-}(L_{1}\oplus L_{2}) 
 & = \mathcal{F}_{\mathrm{b}}(L_{1}\oplus L_{2})\\
 & = \mathcal{F}_{\mathrm{b}}(L_{1})\otimes \mathcal{F}_{\mathrm{b}}(L_{2})\\
 & = [\mathcal{F}_{\mathrm{b}}^{+}(L_{1})\oplus \mathcal{F}_{\mathrm{b}}^{-}(L_{1})]\otimes [\mathcal{F}_{\mathrm{b}}^{+}(L_{2})\oplus \mathcal{F}_{\mathrm{b}}^{-}(L_{2})]\\
 & = [\mathcal{F}_{\mathrm{b}}^{+}(L_{1})\oplus \mathcal{F}_{\mathrm{b}}^{+}(L_{2})]\oplus [\mathcal{F}_{\mathrm{b}}^{-}(L_{1})\oplus \mathcal{F}_{\mathrm{b}}^{-}(L_{2})]\\
 &  \quad \oplus [\mathcal{F}_{\mathrm{b}}^{+}(L_{1})\oplus \mathcal{F}_{\mathrm{b}}^{-}(L_{2})]\oplus [\mathcal{F}_{\mathrm{b}}^{-}(L_{1})\oplus \mathcal{F}_{\mathrm{b}}^{+}(L_{2})].
\end{split}\]

Thus \( U_{+} \) is the unitary operator satisfying \eqref{eq:decomposition_of_Fock_space(+)}.
It is proved similarly that \( U_{-} \) is the unitary operator satisfying
\eqref{eq:decomposition_of_Fock_space(-)}.\end{proof}

Let \( \mathcal{H}_{S}^{\sigma }:=\mathcal{F}_{\mathrm{b}}^{\sigma }(P_{S}(K_{S}\oplus 0)), \)
\( \mathcal{L}_{S}^{\sigma }:=\mathcal{F}_{\mathrm{b}}^{\sigma }(0\oplus E_{S}(\{0\})K_{S}), \)
\( \mathcal{H}_{P_{S}}^{\sigma }:=\mathcal{F}_{\mathrm{b}}^{\sigma }(P_{S}K_{P_{S}}) \)
where \( \sigma =+ \) or \( - \). From the argument to the above, we obtain
the following relations.
\begin{equation}
\label{eq:(H(P_S),+)and(H(P_S),-)}
\mathcal{H}_{P_{S}}^{+}=(\mathcal{H}_{S}^{+}\otimes \mathcal{L}_{S}^{+})\oplus (\mathcal{H}_{S}^{-}\otimes \mathcal{L}_{S}^{-}),\quad \mathcal{H}_{P_{S}}^{-}=(\mathcal{H}_{S}^{+}\otimes \mathcal{L}_{S}^{-})\oplus (\mathcal{H}_{S}^{-}\otimes \mathcal{L}_{S}^{+}).
\end{equation}

\section{Quasifree Representations of \protect\( sp(\infty )\protect \)}

Let \emph{\( K \)} be a complex vector space and \( \gamma (f,g) \) be a non-degenerate
hermitian form for \( f,g\in K \). Let \( \Gamma  \) be an antilinear involution
satisfying \( \Gamma ^{2}=1 \), \( \gamma (\Gamma f,\Gamma g)=-\gamma (g,f) \).
Then we denote finite rank operators on \( K \) satisfying \( \Gamma H\Gamma =-H \)
and \( H^{\dagger }=H \) by \( sp(\infty ) \). \( H^{\dagger } \) is defined
by \( \gamma (H^{\dagger }f,g)=\gamma (f,Hg) \) for all \( f,g\in K \). By
the non-degeneracy of \( \gamma  \), \( H^{\dagger } \) is well-defined. We
call \( H\in sp(\infty ) \) a \textit{Hamiltonian}. \( sp(\infty ) \) is a
Lie algebra endowed with the Lie bracket \( i[H,H']:=i(HH'-H'H) \).

\begin{lm}\label{thm:Finite-Rank-Operatr}\emph{Let \( K_{1} \) be a finite
dimensional subspace of} \( K \). \emph{Then there exists a \( \Gamma  \)-invariant
finite dimensional subspace \( K^{\#}_{1} \) such that \( K_{1}\subset K^{\#}_{1} \)
and the restriction of \( \gamma  \) to \( K^{\#}_{1} \) is non-degenerate}.
\end{lm}
\begin{proof} By the existence of the basis \( \{e_{j}\}_{j=1}^{2k} \) of \( K_{1}^{\#} \)
satisfying 
\begin{gather*}
  \Gamma e_{j}=e_{j},\quad j=1,2,\ldots ,2k,  \\
  \gamma (e_{2j-1},e_{2j})=i,\quad j=1,2,\ldots ,k,  \\
  \gamma (e_{l},e_{l'})=0,\, \, (l,l')\neq (2j-1,2j),(2j,2j-1),j=1,2,\ldots ,k, 
\end{gather*}
this lemma is proved.(See Lemma 4.1 of \cite{Araki-CCR-2}.) \end{proof}

\begin{lm}\label{thm:Hamiltonian} \textit{For any Hamiltonian \( H\in sp(\infty ) \)}\textit{\emph{,}}
\textit{there exists \( f_{j},g_{j}\in (HK)^{\#} \) such that \( Hf=\sum _{j=1}^{N}\gamma (g_{j},f)f_{j} \)
for all \( f\in K \).}\end{lm}

Lemma \ref{thm:Hamiltonian} is verified immediately by using linearly independent
vectors \( \{e_{j}\}_{j=1}^{2k} \) of \( (HK)^{\#} \) given by the proof of
Lemma \ref{thm:Finite-Rank-Operatr}. In fact
\[
Hf=i\sum _{j=1}^{2k}\{\gamma (He_{2j},f)e_{2j-1}-\gamma (He_{2j-1},f)e_{2j}\}\]
 for all \( f\in K \).

For any \( H\in sp(\infty ) \), we can define a second quantization of Hamiltonian,
called a \textit{bilinear Hamiltonian}, \( q(H) \) as follows.

\begin{defini} \textit{\( H\in sp(\infty ) \) satisfies \( Hf=\sum _{j=1}^{N}\gamma (g_{j},f)f_{j} \)
for all \( f\in K \). Then \( q(H):=\frac{1}{2}\sum _{j=1}^{N}B(f_{j})B(g_{j})^{*} \)}.\end{defini} 

Note that the choice of \( f_{j},g_{j} \) is not unique for \( H\in sp(\infty ) \).
However, \( q(H) \) is independent of the choice of \( f_{j},g_{j} \), only
depends on \( H\in sp(\infty ) \)(Lemma 4.4 of \cite{Araki-CCR-2}).

\( q \) is the map from Hamiltonians to bilinear Hamiltonians and satisfies
\( i[q(H),q(H')]=q(i[H,H']) \). Thus \( q_{S}:=\pi _{S}\circ q \) is a representation
of \( sp(\infty ) \) on \( \mathcal{H}_{S} \). \( \widehat{q} \), the map
from Hamiltonians on \( K_{S}\oplus K_{S} \) to bilinear Hamiltonians in \( \mathfrak {A}(K_{S}\oplus K_{S},\widehat{\gamma }_{S},\widehat{\Gamma }_{S}) \),
is defined as well as \( q:sp(\infty )\rightarrow \mathfrak {A}(K,\gamma ,\Gamma ) \).
Let \( \widehat{q}_{P_{S}}:=\pi _{P_{S}}\circ \widehat{q} \), \( q_{P_{S}}(H):=\widehat{q}_{P_{S}}(H\oplus 0) \),
then \( q_{P_{S}} \) is a representation of \( sp(\infty ) \) on \( \mathcal{H}_{P_{S}} \).

\begin{defini}

\begin{enumerate}
\item \emph{A} {*}-\emph{representation \( (\mathcal{H},\pi ) \) of \( sp(\infty ) \)
is called a regular representation if the following two conditions hold} :

\begin{enumerate}
\item \emph{\( i[\pi (H),\pi (H')]=\pi (i[H,H']) \) on a dense set \( \mathcal{H}_{0} \)
of Hilbert space \( \mathcal{H} \) for all \( H,H'\in sp(\infty ) \)}.
\item \emph{\( \pi (H) \) is an essentially self-adjoint operator on \( \mathcal{H}_{0} \)
for all} \( H\in sp(\infty ) \).
\end{enumerate}
\item \emph{Let \( (\mathcal{H}_{j},\pi _{j}) \), \( j=1,2 \) be regular representations
of \( sp(\infty ) \) and \( \mathcal{M}_{j} \) denotes a von Neumann algebra
generated by \( \exp (i\overline{\pi _{j}(H)}),H\in sp(\infty ) \)}. \emph{Then
two representations}, \emph{\( (\mathcal{H}_{1},\pi _{1}) \) and \( (\mathcal{H}_{2},\pi _{2}) \)},
\emph{are quasi-equivalent if there exists a} {*}-\emph{isomorphism of von Neumann
algebras \( \iota :\mathcal{M}_{1}\rightarrow \mathcal{M}_{2} \) such that
\( \iota (\exp (i\overline{\pi _{1}(H))})=\exp (i\overline{\pi _{2}(H)}) \)
for all \( H\in sp(\infty ) \)}. \emph{Then we write \( \pi _{1}\sim _{q}\pi _{2} \)}.
\emph{Moreover}, \emph{if} \( \pi _{1} \) \emph{and \( \pi _{2} \) are unitary
equivalent}, \emph{then we write \( \pi _{1}\sim \pi _{2} \) simply}. 
\end{enumerate}
\end{defini} 

Since all \( q_{S}(H) \) are essentially self-adjoint operators on \( D(\pi _{S}):=\pi _{S}(\mathfrak {A}(K,\gamma ,\Gamma ))\Omega _{S} \),
\( (\mathcal{H}_{S},q_{S}) \) is a regular representation of \( sp(\infty ) \).
Similarly, since all \( q_{P_{S}}(H) \) are essentially self-adjoint operators
on \( D(\pi _{P_{S}}):=\pi _{P_{S}}(\mathfrak {A}(K_{S}\oplus K_{S},\widehat{\gamma }_{S},\widehat{\Gamma }_{S}))\Omega _{P_{S}} \),
\( (\mathcal{H}_{P_{S}},q_{P_{S}}) \) is a regular representation of of \( sp(\infty ) \).
Let 
\[
Q_{S}(H):=\exp \left( i\overline{q_{S}(H)}\right) ,\, \, \, \widehat{Q}_{P_{S}}(\widehat{H}):=\exp \left( i\overline{\widehat{q}_{P_{S}}(\widehat{H})}\right) ,\, \, \, Q_{P_{S}}(H):=\widehat{Q}_{P_{S}}(H\oplus 0)\]
and 
\[
\mathcal{M}_{S}:=\{Q_{S}(H)\, |\, H\in sp(\infty )\}'',\quad \mathcal{M}_{P_{S}}:=\{Q_{P_{S}}(H)\, |\, H\in sp(\infty )\}''.\]

\begin{lm}\label{thm:Q(P_S)=Q(S)x1} \( Q_{P_{S}}(H)=Q_{S}(H)\otimes 1_{\mathcal{L}_{S}} \)
\emph{where} \( 1_{\mathcal{L}_{S}} \) \emph{is the identity operator on} \( \mathcal{L}_{S} \).\end{lm}
\begin{proof} Due to Remark \ref{thm:Remark}, we obtain
\begin{equation}
\label{eq:uQ_S-equal-Q(P_S)u}
u_{\alpha }Q_{S}(H)=Q_{P_{S}}(H)u_{\alpha }
\end{equation}
for all \( H\in sp(\infty ) \). By \eqref{eq:H(P_S)_equal_H(S)xL(S)} and \eqref{eq:uQ_S-equal-Q(P_S)u},
this lemma has been verified.\end{proof}

\emph{Remark}. The following relation is verified quite similar to Lemma \ref{thm:Q(P_S)=Q(S)x1}
: 
\begin{equation}
\label{eq:decomposition_of_W_P(S)(f)}
W_{P_{S}}(f\oplus 0)=W_{S}(f)\otimes 1_{\mathcal{L}_{S}}
\end{equation}
where \( W_{S}(f):=W_{P_{S}}(f\oplus 0)|\mathcal{H}_{S} \) for all \( f\in \mathrm{Re}K_{S} \)
and \( 1_{\mathcal{L}_{S}} \) is the identity operator on \( \mathcal{L}_{S} \). 

In Lemma 5.1 of \cite{Araki-CCR-2}, the following relation has been proved
: If \( P\) is a basis projection on \( K\), then
\begin{equation}
\label{eq:Q(H)W(f)Q(H)*--projective_case}
Q_{P}(H)W_{P}(f)Q_{P}(H)^{*}=W_{P}(e^{iH}f)
\end{equation}
for all \( f\in \mathrm{Re}K \) and \( H\in sp(\infty ) \). Due to Lemma \ref{thm:Q(P_S)=Q(S)x1}
and \eqref{eq:decomposition_of_W_P(S)(f)}, the relation \eqref{eq:Q(H)W(f)Q(H)*--projective_case}
implies the following result : For the general \( S \), we have
\begin{equation}
\label{eq:Q(H)W(f)Q(H)*--non_projective_case}
Q_{S}(H)W_{S}(f)Q_{S}(H)^{*}=W_{S}(e^{iH}f)
\end{equation}
for all \( f\in \mathrm{Re}K \) and \( H\in sp(\infty ) \).

\bigskip

Since \( Q_{S}(H)\mathcal{H}_{S}^{\sigma }\subset \mathcal{H}_{S}^{\sigma } \)
and \( Q_{P_{S}}(H)\mathcal{H}_{P_{S}}^{\sigma }\subset \mathcal{H}_{P_{S}}^{\sigma } \)
where \( \sigma =+ \) or \( - \), 
\[
Q_{S}^{\sigma }(H):=Q_{S}(H)|\mathcal{H}_{S}^{\sigma },\quad Q^{\sigma }_{P_{S}}(H):=Q_{P_{S}}(H)|\mathcal{H}_{P_{S}}^{\sigma }\]
are bounded operators on \( \mathcal{H}_{S}^{\sigma } \) and \( \mathcal{H}_{P_{S}}^{\sigma } \).
We denote the restriction of \( q_{S} \) (resp. \( q_{P_{S}} \)) to \( \mathcal{H}_{S}^{\sigma } \)
(resp. \( \mathcal{H}_{P_{S}}^{\sigma } \)) by \( q_{S}^{\sigma } \) (resp.
\( q_{P_{S}}^{\sigma } \)) and let
\[
\mathcal{M}_{S}^{\sigma }:=\{Q_{S}^{\sigma }(H)\, |\, H\in sp(\infty )\}'',\quad \mathcal{M}_{P_{S}}^{\sigma }:=\{Q_{P_{S}}^{\sigma }(H)\, |\, H\in sp(\infty )\}''.\]

Let \( \mathcal{H}_{1},\mathcal{H}_{2} \) be Hilbert spaces. Then a bounded
operator \( A \) on \( \mathcal{H}_{1}\oplus \mathcal{H}_{2} \) is written
in the form of matrix like this : 
\[
A=\left( \begin{array}{cc}
A_{11} & A_{12}\\
A_{21} & A_{22}
\end{array}\right) ,\, A_{ij}:\mathcal{H}_{j}\rightarrow \mathcal{H}_{i}.\]

We can verify the next lemma immediately.

\begin{lm} \emph{For all} \( H\in sp(\infty ) \), 
\begin{gather*}
 Q_{P_{S}}^{+}(H)=\left( \begin{array}{cc}
Q_{S}^{+}(H)\otimes 1_{+} & 0\\
0 & Q_{S}^{-}(H)\otimes 1_{-}
\end{array}\right), \\
 Q_{P_{S}}^{-}(H)=\left( \begin{array}{cc}
Q_{S}^{+}(H)\otimes 1_{-} & 0\\
0 & Q_{S}^{-}(H)\otimes 1_{+}
\end{array}\right)  
\end{gather*}
\emph{where} \( 1_{\sigma } \) \emph{is the identity operator on} \( \mathcal{L}_{S}^{\sigma } \)
.\end{lm}

If \( S \) is a basis projection, we call \( q_{S}^{\sigma } \) a Fock representation
of \( sp(\infty ) \) on \( \mathcal{H}_{S}^{\sigma } \).

\section{Structure of Fock Representations of \protect\( sp(\infty )\protect \)}

In this section, we assume that \( S \) is a basis projection.

\begin{lm} \emph{Assume that \( K_{S}=K \). Let \( \mathcal{E}_{S} \) be the
set of all C.O.N.S. of \( SK \)}.

\begin{enumerate}
\item \emph{Fix \( e=\{e_{n}\}_{n\in \mathbf{N}}\in \mathcal{E}_{S} \)}. \emph{Then
\( \{e_{n},\Gamma e_{n}\}_{n\in \mathbf{N}} \) is a C.O.N.S. of \( K \)}.
\item \emph{For \( g,h\in K \)}, 
\[
H_{gh}f:=\gamma (g,f)h+\gamma (h,f)g+\gamma (\Gamma g,f)\Gamma h+\gamma (\Gamma h,f)\Gamma g,\quad f\in K.\]
\( H_{gh} \) \emph{satisfies} \( \Gamma H_{gh}\Gamma =-H_{gh} \), \( H_{gh}^{\dagger }=H_{gh} \).
\emph{Let
\begin{gather*}
H(e;1,k,l):=H_{e_{k},e_{l}}, \quad H(e;2,k,l):=H_{e_{k},\Gamma e_{l}},\\
H(e;3,k,l):=H_{ie_{k},e_{l}}, \quad H(e;4,k,l):=H_{ie_{k},\Gamma e_{l}}.
\end{gather*}
Then \( sp(\infty ) \) is real-linearly spanned by \( H(e;j,k,l) \), \( 1\leq j\leq 4 \),
\( k,l\in \mathbf{N} \)}, \( e\in \mathcal{E}_{S} \). \emph{Moreover},
\begin{gather}
 q(H(e;1,k,l))=B(e_{k})B(e_{l})^{*}+B(e_{l})B(e_{k})^{*}+\delta _{kl}, \label{eq:q(H(e;1,k,l))} \\
 q(H(e;2,k,l))=B(e_{k})^{*}B(e_{l})^{*}+B(e_{k})B(e_{l}), \label{eq:q(H(e;2,k,l))} \\
 q(H(e;3,k,l))=i[B(e_{k})B(e_{l})^{*}-B(e_{l})B(e_{k})^{*}], \label{eq:q(H(e;3,k,l))} \\
 q(H(e;4,k,l))=i[B(e_{k})B(e_{l})-B(e_{k})^{*}B(e_{l})^{*}], \label{eq:q(H(e;4,k,l))} 
\end{gather}
 
\end{enumerate}
\end{lm}
\begin{proof} (1) obvious. (2) For all \( H\in sp(\infty ) \), there
exist \( f_{j},g_{j}\in (HK)^{\#} \)
such that \( Hf=\sum _{j=1}^{N}\gamma (g_{j},f)f_{j} \). Let \( \{e_{n}\}_{n=1}^{M} \)
be a C.O.N.S. of \( S(HK)^{\#} \). Then
\[
x=\sum _{n=1}^{M}(e_{n},x)_{S}e_{n}+(\Gamma e_{n},x)_{S}\Gamma e_{n}=\sum _{n=1}^{M}\gamma (e_{n},x)e_{n}-\gamma (\Gamma e_{n},x)\Gamma e_{n}\]
for all \( x\in (HK)^{\#} \). In particular, let \( x=f_{j},g_{j} \), then
we have
\[\begin{split}
Hf & = \sum _{j=1}^{N}\gamma (g_{j},f)f_{j}\\
   & = \sum _{k,l=1}^{M}\{\alpha (1,k,l)\gamma (e_{k},f)e_{l}-\alpha (2,k,l)\gamma (\Gamma e_{k},f)e_{l}\\
   & \quad +\alpha (3,k,l)\gamma (\Gamma e_{k},f)\Gamma e_{l}-\alpha (4,k,l)\gamma (e_{k},f)\Gamma e_{l}\}.
\end{split}\]
\( \alpha (j,k,l) \) is defined as follows :
\begin{eqnarray*}
 & \alpha (1,k,l):=\sum _{j=1}^{N}\overline{\gamma
(e_{k},g_{j})}\gamma (e_{l},f_{j}), & \\
 & \alpha (2,k,l):=\sum _{j=1}^{N}\overline{\gamma (\Gamma
e_{k},g_{j})}\gamma (e_{l},f_{j}), & \\
 & \alpha (3,k,l):=\sum _{j=1}^{N}\overline{\gamma (\Gamma
e_{k},g_{j})}\gamma (\Gamma e_{l},f_{j}), & \\
 & \alpha (4,k,l):=\sum _{j=1}^{N}\overline{\gamma (e_{k},g_{j})}\gamma (\Gamma e_{l},f_{j}).& 
\end{eqnarray*}
By \( \gamma (Hf,g)=\gamma (f,Hg) \) for all \( f,g\in K \), \( \alpha (j,k,l) \)
satisfies
\begin{gather}
 \overline{\alpha (4,k,l)}=\alpha (2,l,k), \label{eq:relation_of_alpha(j,k,l)_No1} \\
 \overline{\alpha (1,k,l)}=\alpha (1,l,k), \label{eq:relation_of_alpha(j,k,l)_No2} \\
 \overline{\alpha (3,k,l)}=\alpha (3,l,k). \label{eq:relation_of_alpha(j,k,l)_No3} 
\end{gather}
In fact, if \( f=e_{k} \) and \( g=\Gamma e_{l} \), then we have \eqref{eq:relation_of_alpha(j,k,l)_No1},
if \( f=e_{k} \) and \( g=e_{l} \), we have \eqref{eq:relation_of_alpha(j,k,l)_No2},
if \( f=\Gamma e_{k} \) and \( g=\Gamma e_{l} \), we have \eqref{eq:relation_of_alpha(j,k,l)_No3}.

On the other hand, by \( \Gamma H\Gamma =-H \), \( \alpha (i,j,k) \) satisfies
\begin{gather}
 \overline{\alpha (3,k,l)}=\alpha (1,k,l), \label{eq:relation_of_alpha(j,k,l)_No4} \\
 \overline{\alpha (4,k,l)}=\alpha (2,k,l). \label{eq:relation_of_alpha(j,k,l)_No5} 
\end{gather}
Now let \( e \) denote a C.O.N.S. of \( SK \) satisfying \( e\supset \{e_{n}\}_{n=1}^{M} \).
By \eqref{eq:relation_of_alpha(j,k,l)_No1} \( - \) \eqref{eq:relation_of_alpha(j,k,l)_No5},
we have 
\begin{gather*}
 H=\sum _{1\leq k<l\leq M}H(e;k,l)+\frac{1}{2}\sum _{1\leq k\leq M}H(e;k,k), \\
 \begin{split}
 H(e;k,l):&=\mathrm{Re}[\alpha (1,k,l)]H(e;1,k,l)-\mathrm{Im}[\alpha (1,k,l)]H(e;3,k,l)\\
 & \quad -\mathrm{Re}[\alpha (2,k,l)]H(e;2,k,l)-\mathrm{Im}[\alpha
(2,k,l)]H(e;4,k,l).
\end{split} 
\end{gather*}
Thus \( sp(\infty ) \) is real-linearly spanned by all \( H(e;j,k,l) \).\end{proof}

\begin{lm}\label{thm:cyclic_vector} 

\begin{enumerate}
\item \( \Omega _{S} \) \emph{is a cyclic vector for} \( \mathcal{M}_{S}^{+} \).
\item \emph{For any} \( e_{1}\in SK \), \( \pi _{S}(B(e_{1}))\Omega _{S} \) \emph{is
a cyclic vector for} \( \mathcal{M}_{S}^{-} \)\emph{. }
\end{enumerate}
\end{lm}
\begin{proof} (1) Fix \( e=\{e_{n}\}_{n=1}^{\infty }\in \mathcal{E}_{S} \).
By \eqref{eq:q(H(e;2,k,l))} and \eqref{eq:q(H(e;4,k,l))}, we have 
\[
q_{S}(H(e;2,k,l))+i^{-1}q_{S}(H(e;4,k,l))=2\pi _{S}(B(e_{k})B(e_{l}))\]
 for all \( k,l \). Since \( \mathcal{H}_{S}^{+} \) is linearly spanned by
elements \( \prod _{k,l}\pi _{S}(B(e_{k})B(e_{l}))\Omega _{S} \), \( \Omega _{S} \)
is a cyclic vector for \( \mathcal{M}_{S}^{+} \).

(2) Fix \( e:=\{e_{n}\}_{n=1}^{\infty }\in \mathcal{E}_{S} \). By \eqref{eq:q(H(e;1,k,l))}
and \eqref{eq:q(H(e;3,k,l))}, we have 
\[
q_{S}(H(e;1,k,l))+i^{-1}q_{S}(H(e;3,k,l))=2\pi _{S}(B(e_{k})B(e_{l})^{*}),\, k\neq l.\]
Since 
\[
\{q_{S}(H(e;1,k,1))+i^{-1}q_{S}(H(e;3,k,1))\}\pi _{S}(B(e_{1}))\Omega _{S}=\pi _{S}(B(e_{k}))\Omega _{S}\, (k\geq 2),\]
\( SK \) is generated by \( \pi _{S}(B(e_{1}))\Omega _{S} \) and all \( q_{S}(H(e;1,k,1)) \),
\( q_{S}(H(e;3,k,1)) \). Therefore \( \pi _{S}(B(e_{1}))\Omega _{S} \) is
a cyclic vector for \( \mathcal{M}_{S}^{-} \).\end{proof}

\begin{lm}\label{thm:number_operator}

\begin{enumerate}
\item \emph{Let 
\begin{gather*}
 D(N_{S}):=\left\{ \xi =\oplus _{n=0}^{\infty }\xi ^{(n)}\in \mathcal{H}_{S}\, \Big|\, \xi ^{(n)}\in \otimes _{\mathrm{s}}^{n}SK_{S},\, \sum _{n=1}^{\infty }n^{2}\parallel \xi ^{(n)}\parallel ^{2}<\infty \right\} ,\\
 (N_{S}\xi )^{(n)}:=n\xi ^{(n)},\quad \xi \in D(N_{S}).  
\end{gather*}
Then the number operator \( N_{S} \) is an essentially self-adjoint operator
on the finite particle vector subspace of} \( \mathcal{H}_{S} \).
\item \( \sigma (N_{S})=\sigma _{\mathrm{p}}(N_{S})=\mathbf{N}\cup \{0\} \).
\item \( e^{i\overline{N_{S}}}\in \mathcal{M}_{S} \).
\end{enumerate}
\end{lm} 
\begin{proof} (1) and (2) are well-known facts. 

(3) Fix \( e=\{e_{n}\}\in \mathcal{E}_{S} \). For \( \Lambda \subset \mathbf{N} \),
\( \#\Lambda <\infty  \), we define \( N_{\Lambda }:=\sum _{j\in \Lambda }\pi _{S}(B(e_{j})B(e_{j})^{*}) \).
Since \( N_{\Lambda }+\frac{1}{2}\#\Lambda =\frac{1}{2}\sum _{k\in \Lambda }q_{S}(H(e;1,k,k)) \),
\( N_{\Lambda }+\frac{1}{2}\#\Lambda  \) is an essentially self-adjoint operator
on the finite particle vector subspace of \( \mathcal{H}_{S} \) and \( \exp (i\overline{N_{\Lambda }})\in \mathcal{M}_{S} \).
Since 
\[
N_{\Lambda }\pi _{S}(B(e_{\lambda _{1}})\ldots B(e_{\lambda_{n}}))\Omega _{S}
 =\sum _{j\in \Lambda }\#\{k\, |\, e_{\lambda _{k}}=e_{j}\}\pi _{S}(B(e_{\lambda _{1}})\ldots B(e_{\lambda _{n}}))\Omega _{S},\]
we have
\[\begin{split}
\exp & (i\overline{N_{\Lambda }})\pi _{S}(B(e_{\lambda _{1}})\ldots
B(e_{\lambda _{n}}))\Omega _{S}\\
& =\exp \left( i\sum _{j\in \Lambda }\#\{k\, |\, e_{\lambda _{k}}=e_{j}\}\right)\pi _{S}(B(e_{\lambda _{1}})\ldots B(e_{\lambda _{n}}) \Omega _{S}.\end{split}\]
If \( \Lambda \rightarrow \mathbf{N}\cup \{0\} \), then \( \sum _{j\in \Lambda }\#\{k\, |\, e_{\lambda _{k}}=e_{j}\}\rightarrow n \).
Thus
\[\begin{split}
\textrm{s}\frac{\, \, }{\, \, }\lim _{\Lambda \rightarrow
\mathbf{N}\cup \{0\}}\exp (i\overline{N_{\Lambda }})\pi
_{S}(B(e_{\lambda _{1}})\ldots B(e_{\lambda _{n}}))\Omega _{S}
 & =\exp (in)\pi _{S}(B(e_{\lambda _{1}})\ldots B(e_{\lambda _{n}}))\Omega _{S}\\
 & = \exp (i\overline{N_{S}})\pi _{S}(B(e_{\lambda _{1}})\ldots B(e_{\lambda _{n}}))\Omega _{S}.
\end{split}\]
This shows that \( \exp (i\overline{N_{S}})=\mathrm{s}\frac{\, \, }{\, \, }\lim _{\Lambda \rightarrow \mathbf{N}\cup \{0\}}\exp (i\overline{N_{\Lambda }})\in \mathcal{M}_{S} \).\end{proof}

\begin{lm} \( \mathcal{M}_{S}^{\sigma } \) \emph{acts irreducibly on} \( \mathcal{H}_{S}^{\sigma } \).\end{lm}
\begin{proof} Assume that \( \mathcal{H}_{0} \) is a non-zero intersection
of \( \mathcal{M}_{S}^{+} \)-invariant closed subspace of \( \mathcal{H}_{S}^{+} \)
and the finite particle vector subspace of \( \mathcal{H}_{S}^{+}
\). The spectrum \( \sigma (N_S|\mathcal{H}_{0})\) is contained in \(
\sigma (N_{S}|\mathcal{H}_{S}^{+})=2\mathbf{N} \cup \{ 0\} \) due to the
invariance of \( \mathcal{H}_0\) under \( \mathcal{M}_{S}^{+} \). 
Let \( \lambda _{0}:=\min \sigma (N_{S}|\mathcal{H}_{0}) \).

If \( \lambda _{0}=0 \), we have \( \Omega _{S}\in \mathcal{H}_{0} \). Thus
\( \overline{\mathcal{H}_{0}}\supset \mathcal{M}_{S}^{+}\overline{\mathcal{H}_{0}}\supset \mathcal{M}_{S}^{+}\Omega _{S}=\mathcal{H}_{S}^{+} \).
This shows that \( \mathcal{M}_{S}^{+} \) acts irreducibly on \( \mathcal{H}_{S}^{+} \).

If \( \lambda _{0}=2 \), there exists \( \xi \in \mathcal{H}_{0} \) such that
\( N_{S}\xi =2\xi  \), \( \xi \in \otimes ^{2}_{\mathrm{s}}SK \). Then we
can show that there exist \( j,k\in \mathbf{N} \) such that \( \pi
_{S}(B(e_{j})^{*})\pi _{S}(B(e_{k})^{*})\xi \) is a non-zero element
of \( \mathcal{H}_0\).
Since
\[
\pi _{S}(B(e_{j})^{*}B(e_{k})^{*})N_{S}=N_{S}\pi _{S}(B(e_{j})^{*}B(e_{k})^{*})+2\pi _{S}(B(e_{j})^{*}B(e_{k})^{*}),\]
we have
\[\begin{split}
N_{S}\pi _{S}(B(e_{j})^{*}B(e_{k})^{*})\xi & = \{\pi _{S}(B(e_{j})^{*}B(e_{k})^{*})N_{S}-2\pi _{S}(B(e_{j})^{*}B(e_{k})^{*})\}\xi \\
 & = \pi _{S}(B(e_{j})^{*}B(e_{k})^{*})\cdot 2\xi -2\pi_{S}(B(e_{j})^{*}B(e_{k})^{*}\xi\\
 & =0.
\end{split}\]
This shows \( \pi _{S}(B(e_{j})^{*}B(e_{k})^{*})\xi \in \mathbf{C}\Omega _{S} \)
and we have \( \lambda _{0}=0 \). This is contradiction. Thus \( \lambda _{0}\neq 2 \).
Moreover, \( \lambda _{0}\neq 2n,\, n=1,2,\ldots  \) is proved by induction. 

The case of \( \mathcal{M}_{S}^{-} \) is quite similar.\end{proof}

\begin{lm} \emph{Two representations} \( q_{S}^{+} \) \emph{and} \( q_{S}^{-} \)
\emph{are not quasi-equivalent}.\end{lm}
\begin{proof} By the irreducibility of \( \mathcal{M}_{S}^{\sigma } \), it
suffices that \( q_{S}^{+} \) and \( q_{S}^{-} \) are not unitary equivalent.
Suppose that there exists an unitary operator \( V:\mathcal{H}_{S}^{+}\rightarrow \mathcal{H}_{S}^{-} \)
such that \( VQ_{S}^{+}(H)=Q_{S}^{-}(H)V \) for all \( H\in sp(\infty ) \).
Since the left hand side of \( VQ_{S}^{+}(tH)\Omega _{S}=Q_{S}^{-}(tH)V\Omega _{S},\, t\in \mathbf{R} \)
is differentiable at \( t\in \mathbf{R} \), the differential of the right hand
side make sense and the domain of \( \overline{q_{S}^{-}(H)} \) contains \( V\Omega _{S} \).
Now let \( H=H(e;1,k,l) \) and \( H(e;3,k,l) \), then we have \( \overline{\pi _{S}(B(e_{k})B(e_{l})^{*})}V\Omega _{S}=V\pi _{S}(B(e_{k})B(e_{l})^{*})\Omega _{S}=0 \)
for all \( k,l \). Therefore, since \( \overline{\pi _{S}(B(e_{l})^{*})}V\Omega _{S}=0 \)
for all \( l \), we obtain \( V\Omega _{S}=0 \). This is contradiction.\end{proof}

\section{Structure of non-Fock Representations of \protect\( sp(\infty )\protect \)}

In this section, we assume that \( S \) is not a basis projection. 

Lemma \ref{thm:von_Neumann_algebra_rep==subrep} is a well-known fact.

\begin{lm}\label{thm:von_Neumann_algebra_rep==subrep} \emph{Let} \( \mathcal{M} \)
\emph{be a von Neumann algebra on a Hilbert space of} \( \mathcal{H} \) \emph{and}
\( E\in \mathcal{M}' \) \emph{be a projection}. \emph{Let} \( C(\mathcal{M}):=\mathcal{M}\cap \mathcal{M}' \)
\emph{and} \( \mathcal{M}_{E}=\{Q_{E}\, |\, Q\in \mathcal{M}\} \), \( Q_{E}:=Q|E\mathcal{H} \).
\emph{Then the map} \( \iota :\mathcal{M}\ni Q\mapsto Q_{E}\in \mathcal{M}_{E} \)
\emph{is a} {*}-\emph{homomorphism and continuous with respect to the strong
operator topologies}. \emph{Moreover} \( \iota  \) \emph{is a} {*}-\emph{isomorphism
if and only if} \( C(E):=\min \{F\in C(\mathcal{M})\, |\, F\geq E,F^{2}=F^{*}=F\}=1 \).\end{lm}

\begin{lm}\label{thm:S_not-projection=q_S==q_S(+)==q_S(-)} \emph{Let}
\( E_{+} \) \emph{be a projection of} \( \mathcal{H}_{S} \) \emph{on} \( \mathcal{H}_{S}^{+} \).
(\( E_{+} \) \emph{is a element of} \( \mathcal{M}_{S}'.) \) \emph{Then} \( C(E_{+})=1 \).
\emph{Therefore}, \( q_{S} \), \( q_{S}^{+} \) \emph{and} \( q_{S}^{-} \)
\emph{are all quasi-equivalent}.\end{lm}
\begin{proof} By Lemma \ref{thm:von_Neumann_algebra_rep==subrep}, we prove
that \( C(E_{+}) \) is the identity operator on \( \mathcal{H}_{S} \). Since
\( C(E_{+})|\mathcal{H}_{S}^{+} \) is the identity operator on \( \mathcal{H}_{S}^{+} \),
we have only to show that \( C(E_{+})|\mathcal{H}_{S}^{-} \) is the identity
operator on \( \mathcal{H}_{S}^{-} \). We claim that 
\begin{equation}
\label{eq:C(E(+))B(g+0)}
C(E_{+})\pi _{P_{S}}(B(g\oplus 0))\Omega _{P_{S}}=\pi _{P_{S}}(B(g\oplus 0))\Omega _{P_{S}}
\end{equation}
 for all \( g\in K_{S} \).

(i) The case of \( g\in E_{S}((0,1))K_{S} \).

Let \( g':=\sqrt{(1-S)^{-1}S}g \) for \( g\in E_{S}((0,1))K_{S}\cap D((1-S)^{-1}) \).
(\( D(A) \) is the domain of operator \( A \).) \( g' \) is an element of
\( E_{S}((0,1))K_{S} \) and satisfies \( P_{S}(g\oplus 0)=P_{S}(0\oplus g') \)
and \( \pi _{P_{S}}(B(g\oplus 0))\Omega _{P_{S}}=\pi _{P_{S}}(B(0\oplus g'))\Omega _{P_{S}} \).
Thus it suffices that
\begin{equation}
\label{eq:C(E(+))B(0+g)}
C(E_{+})\pi _{P_{S}}(B(0\oplus g'))\Omega _{P_{S}}=\pi _{P_{S}}(B(0\oplus g'))\Omega _{P_{S}}
\end{equation}
for all \( g'\in E_{S}((0,1))K_{S} \). Now since \( [q(H\oplus 0),B(0\oplus g')]=0 \)
for all \( H\in sp(\infty ) \), \( C(E_{+})\in \mathcal{M}_{S} \) ``\emph{commutes}''
with \( \pi _{P_{S}}(B(0\oplus g'))|\mathcal{H}_{S}\cap D(\pi _{P_{S}}) \).
(If a bounded operator \( Q \) on a Hilbert space \( \mathcal{H} \) and (unbounded)
operator \( A \) on a Hilbert space \( \mathcal{H} \) with a dense domain
\( D(A) \) satisfy \( \left\langle QA\eta ,\xi \right\rangle =\left\langle Q\eta ,A^{*}\xi \right\rangle  \)
for all \( \eta ,\xi \in D(A) \), it is said that \( A \) commutes with \( Q \).)
Thus we obtain \eqref{eq:C(E(+))B(0+g)}.

(ii) The case of \( g\in E_{S}(\{1\})K_{S} \).

Let \( f \) be an unit vector in \( E_{S}((0,1))K_{S} \). Then
\[ \begin{split}
\pi _{P_S}& (B(g\oplus 0))\Omega _{P_S}\\
&=\pi _{P_S}(B(g\oplus 0))\pi _{P_S}(B(P_S(f\oplus 0))^*B(P_S(f\oplus 0)))\Omega _{P_S}\\
&=\pi _{P_S}(B(g\oplus 0))\pi _{P_S}(B(P_S(f\oplus 0))^*B(f\oplus 0))\Omega _{P_S}\\
&=\pi _{P_S}(B(g\oplus 0)B(\Gamma _S \gamma _S^{-1}Sf\oplus 0)\pi _{P_S}(B(f\oplus 0))\Omega _{P_S}\\
&\quad +\pi _{P_S}(B(g\oplus 0)B(f\oplus 0))\pi _{P_S}(B(0\oplus 
\Gamma _S(-\gamma _S^{-1}) 2\sqrt{S(1-S)} f))\Omega _{P_S}.
\end{split}\]
Since \( \pi _{P_{S}}(B(g\oplus 0)B(\Gamma _{S}\gamma _{S}^{-1}Sf\oplus 0))|\mathcal{H}_{S}\cap D(\pi _{P_{S}}) \)
and \( \pi _{S}(B(g\oplus 0)B(f\oplus 0))|\mathcal{H}_{S}\cap D(\pi _{P_{S}}) \)
commute with \( C(E_{+}) \) and \( \Gamma _{S}(-\gamma _{S}^{-1})\sqrt{S(1-S)}f \)
is an element of \( E_{S}((0,1))K_{S} \), we can use the case (i) and obtain
\eqref{eq:C(E(+))B(g+0)}.\end{proof}

\begin{cor}\label{thm:S:non-projection==>q_S=q_P(S)} \( q_{S}^{\sigma } \)
\emph{and} \( q_{P_{S}}^{\sigma } \) \emph{are quasi-equivalent}.\end{cor}
\begin{proof} \( \Omega _{P_{S}} \) is a separating vector for \( C(\mathcal{M}_{P_{S}}^{+}) \).
Indeed, let 
\[
\left( \begin{array}{cc}
a_{+}\otimes 1_{+} & 0\\
0 & a_{-}\otimes 1_{-}
\end{array}\right) \Omega _{P_{S}}=0,\, a_{\sigma }\in C(\mathcal{M}_{S}^{\sigma }),\]
then \( a_{+}\Omega _{S}=0 \). Since \( \Omega _{S} \) is a separating vector
for \( C(\mathcal{M}_{S}^{+}) \), we have \( a_{+}=0 \). Moreover, by \( q_{S}^{+}\sim _{q}q_{S}^{-} \)(Lemma
\ref{thm:S_not-projection=q_S==q_S(+)==q_S(-)}), we have \( a_{-}=0 \)
as well. 

Let \( \widehat{E}_{+}:\mathcal{H}^{+}_{P_{S}}\rightarrow \mathcal{H}_{S}^{+} \)
be a projection and \( C(\widehat{E}_{+}):=\min \{E\in C(\mathcal{M}_{P_{S}}^{+})\, |\, E^{2}=E^{*}=E,E\geq \widehat{E}_{+}\} \).
Since \( C(\widehat{E}_{+})\Omega _{P_{S}}=\Omega _{P_{S}} \) and \( \Omega _{P_{S}} \)
is a separating vector for \( C(\mathcal{M}_{P_{S}}^{+}) \), we have
\( C(\widehat{E}_{+})=1 \), that is, 
\( q_{S}^{+}\sim _{q}q_{P_{S}}^{+} \). By Lemma \ref{thm:S_not-projection=q_S==q_S(+)==q_S(-)},
we obtain
\[q_{S}^{+}\sim _{q}q_{P_{S}}^{+}\sim _{q}q_{S}^{-}\sim _{q}q_{P_{S}}^{-}.\]
\end{proof}

If \( 0<S<1 \), the commutant \( (\mathcal{M}_{P_{S}}^{+})' \) is written
explicitly by using Tomita-Takesaki theory and we can show that \( \mathcal{M}_{P_{S}}^{+} \)
is a factor.

\begin{lm} \emph{Let} \( 0<S<1 \). \emph{Then} \( \Omega _{P_{S}} \) \emph{is
a cyclic and separating vector for} \( \mathcal{M}_{P_{S}}^{+} \).\end{lm}
\begin{proof} If \( 0<S<1 \), we have already got \( \mathcal{H}_{P_{S}}^{+}=\mathcal{H}_{S}^{+} \)
and \( Q_{P_{S}}(H)=Q_{S}(H) \). Thus \( \Omega _{P_{S}} \) is cyclic for
\( \mathcal{M}_{P_{S}}^{+} \). Since \( \{\widehat{Q}_{P_{S}}(0\oplus H)\Omega _{P_{S}}\, |\, H\in sp(\infty )\} \)
generate \( \mathcal{H}_{P_{S}}^{+} \) and \( \widehat{Q}_{P_{S}}(0\oplus H)\in (\mathcal{M}_{P_{S}}^{+})' \),
\( \Omega _{P_{S}} \) is cyclic for \( (\mathcal{M}_{P_{S}}^{+})' \) i.e.
\( \Omega _{P_{S}} \) is separating for \( \mathcal{M}_{P_{S}}^{+} \).\end{proof}

We define the modular conjugation \( J_{_{\Omega _{P_{S}}}} \) and the modular
operator \( \Delta _{\Omega _{P_{S}}} \) for \( \mathcal{M}_{P_{S}}^{+} \).

For a bijective linear (resp. conjugate linear) operator \( U \) on \( K_{S}\oplus K_{S} \),
we define a {*}-automorphism (resp. conjugate {*}-automorphism) \( \tau (U) \)
of \( \mathfrak {A}(K_{S}\oplus K_{S},\widehat{\gamma }_{S},\widehat{\Gamma }_{S}) \)
satisfying \( \tau (U)B(h)=B(Uh) \). 

Let \( \omega (f\oplus g):=\Gamma _{S}g\oplus \Gamma _{S}f \). The conjugate
linear map \( J_{\Omega _{P_{S}}} \) is defined by 
\[
J_{\Omega _{P_{S}}}\pi _{P_{S}}(A)\Omega _{P_{S}}=\pi _{P_{S}}(\tau (\omega )A)\Omega _{P_{S}},\quad A\in \mathfrak {A}(K_{S}\oplus K_{S},\widehat{\gamma }_{S},\widehat{\Gamma }_{S}).\]

Let \( 0<S<1 \) and \( H_{S}:=\log (S(1-S)^{-1}) \). Let \( \Theta _{S} \)
be an infinitesimal generator defined by
\[
\exp (it\Theta _{S})\pi _{S}(A)\Omega _{S}=\pi _{S}(\tau (e^{itH_{S}})A)\Omega _{S},\quad A\in \mathfrak {A}(K,\gamma ,\Gamma ).\]
\( \alpha  \) is defined in Remark \ref{thm:Remark}. Then \( \Delta _{\Omega _{P_{S}}}:=e^{-\Theta _{S}} \).
Due to \( 0<S<1 \), \( \Delta _{\Omega _{P_{S}}} \) is defined on a dense
set of \( \mathcal{H}_{P_{S}}=\mathcal{H}_{S} \). (See Remark \ref{thm:Remark}.)

\begin{lm} \emph{Let} \( 0<S<1 \).

\begin{enumerate}
\item \emph{The restriction of \( J_{\Omega _{P_{S}}} \) on \( \mathcal{H}_{P_{S}}^{+} \)
is the modular conjugation associated with the pair \( (\mathcal{M}_{P_{S}}^{+},\Omega _{P_{S}}) \)
and the restriction of \( \Delta _{\Omega _{P_{S}}} \)on \( \mathcal{H}_{P_{S}}^{+} \)
is the modular operator}.
\item \( \mathcal{M}_{P_{S}}^{+} \) \emph{is a factor}.
\end{enumerate}
\end{lm}
\begin{proof} (1) We obtain \( J_{\Omega _{P_{S}}}Q_{P_{S}}(H)\Omega _{P_{S}}=\Delta _{\Omega _{P_{S}}}^{1/2}Q_{P_{S}}(H)^{*}\Omega _{P_{S}} \)
for all \( H\in sp(\infty ) \). For any \( A\in \mathcal{M}_{P_{S}}^{+} \),
there exists a net \( \{A_{\nu }\} \) in the linear hull of \( \{Q_{P_{S}}(H)\, |\, H\in sp(\infty )\} \)
such that \( A_{\nu }\rightarrow A \) \( (\nu \to \infty) \) with
respect to the strong {*} operator topology
:
\[
\left\Vert (A_{\nu }-A)x\right\Vert ^{2}+\left\Vert (A^{*}_{\nu }-A^{*})x\right\Vert ^{2}\rightarrow 0\]
for all \( x\in \mathcal{H}_{P_{S}}^{+} \). Therefore we have
\[
\langle \Delta _{P_{S}}^{1/2}\Psi ,A^{*}\Omega _{P_{S}}\rangle =\lim _{\nu \rightarrow \infty }\langle \Delta _{P_{S}}^{1/2}\Psi ,A_{\nu }^{*}\Omega _{P_{S}}\rangle =\lim _{\nu \rightarrow \infty }\langle \Psi ,J_{\Omega _{P_{S}}}A_{\nu }\Omega _{P_{S}}\rangle =\langle \Psi ,J_{\Omega _{P_{S}}}A\Omega _{P_{S}}\rangle \]
 for all \( \Psi  \) in the domain of \( \Delta _{\Omega _{P_{S}}} \). This
shows that all \( A^{*}\Omega _{P_{S}} \) are elements of the domain of \( \Delta _{\Omega _{P_{S}}} \).
Let \( T_{\Omega _{P_{S}}}:=J_{\Omega _{P_{S}}}\Delta _{\Omega _{P_{S}}}^{1/2} \),
then \( T_{\Omega _{P_{S}}}A\Omega _{P_{S}}=A^{*}\Omega _{P_{S}} \) for all
\( A\in \mathcal{M}_{P_{S}}^{+} \). Thus \( J_{\Omega _{P_{S}}} \) is the
modular conjugation associated with the pair \( (\mathcal{M}_{P_{S}}^{+},\Omega _{P_{S}}) \)
and \( \Delta _{\Omega _{P_{S}}} \) is the modular operator. 

(2) \( Q_{P_{S}}(H) \) satisfies \( [W_{P_{S}}(0\oplus f),Q_{P_{S}}(H)]=0 \)
from the direct computation and this shows
\begin{equation}
\label{eq:R(ReK+0)_supset_M(P_S)}
\mathcal{R}_{P_{S}}(\mathrm{Re}K_{S}\oplus 0)=\mathcal{R}_{P_{S}}(0\oplus \mathrm{Re}K_{S})'\supset \mathcal{M}_{P_{S}}=\mathcal{M}_{P_{S}}^{+}\oplus \mathcal{M}_{P_{S}}^{-}.
\end{equation}
By Tomita-Takesaki theory we have 
\begin{equation}
\label{eq:commutant_of_M(P_S)}
(\mathcal{M}_{P_{S}}^{+})'=J_{\Omega _{P_{S}}}\mathcal{M}_{P_{S}}^{+}J_{\Omega _{P_{S}}}=\{\widehat{Q}_{P_{S}}(0\oplus H)|\mathcal{H}_{P_{S}}^{+}\, |\, H\in sp(\infty )\}''.
\end{equation}
From the quasi-equivalence of representations \( q_{P_{S}}^{+} \) 
and \( q_{P_{S}}^{-} \),
\eqref{eq:commutant_of_M(P_S)} implies that \( (\mathcal{M}_{P_{S}}^{-})' \) is
generated by all \( \widehat{Q}_{P_{S}}(0\oplus H)|\mathcal{H}_{P_{S}}^{-} \),
\( H\in sp(\infty ) \). Thus \( (\mathcal{M}_{P_{S}}^{+})'\oplus (\mathcal{M}_{P_{S}}^{-})' \)
is generated by all \( \widehat{Q}_{P_{S}}(0\oplus H) \), \( H\in sp(\infty ) \).
Since \( [W_{P_{S}}(f\oplus 0),\widehat{Q}_{P_{S}}(0\oplus H)]=0 \), we have
\begin{equation}
\label{eq:R'_supset_M(+)'+M(-)'}
\mathcal{R}_{P_{S}}(\mathrm{Re}K_{S}\oplus 0)'=\mathcal{R}_{P_{S}}(0\oplus \mathrm{Re}K_{S})\supset (\mathcal{M}_{P_{S}}^{+})'\oplus (\mathcal{M}_{P_{S}}^{-})'.
\end{equation}
From Corollary \ref{thm:factoriarity_of_R(ReK+0)} and \eqref{eq:R(ReK+0)_supset_M(P_S)}
and \eqref{eq:R'_supset_M(+)'+M(-)'}, we have
\[
C(\mathcal{M}_{P_{S}}^{+})\oplus C(\mathcal{M}_{P_{S}}^{-})\subset C(\mathcal{R}_{P_{S}}(\mathrm{Re}K_{S}\oplus 0))=\mathbf{C}1.\]
Thus \( C(\mathcal{M}_{P_{S}}^{+})=\mathbf{C}1_{+} \).(And \( C(\mathcal{M}_{P_{S}}^{-})=\mathbf{C}1_{-} \).)\end{proof}

\section{Quasi-equivalence of\\ \protect\( \qquad \protect \)Quasifree Representations
of \protect\( sp(\infty )\protect \)}

Let \( \mathfrak {S} \) be the set of all positive semi-definite hermitian
forms \( S \) on \( K \) satisfying \eqref{eq:S(f,g)}. Now we give the main
result of this paper.

\begin{thm}\label{thm:Main_Theorem} \textit{Assume that \( K \) is separable}\textit{\emph{.}}
\textit{Let \( S,\, S'\in \mathfrak {S} \). Two quasifree representations \( q^{\sigma }_{S} \)
and \( q^{\sigma }_{S'} \) of \( sp(\infty ) \) are quasi-equivalent if and
only if the following two conditions hold} \textit{\emph{:}}

\begin{enumerate}
\item \textit{The topologies induced by \( \parallel f\parallel _{S} \) and \( \parallel f\parallel _{S'} \)
on \( K \) are equivalent}\textit{\emph{,}} \textit{i}\textit{\emph{.}}\textit{e}\textit{\emph{.}}
\textit{there exists \( \beta >\alpha >0 \) such that \( \alpha \parallel f\parallel _{S}\leq \parallel f\parallel _{S'}\leq \beta \parallel f\parallel _{S} \)
for all \( f\in K \).}
\item \textit{\( 1-\rho (S)e^{-\chi (S)}e^{\chi (S')}\rho (S') \) is a Hilbert-Schmidt
class operator on \( K_{S} \)} \textit{\emph{}}\textit{where \( \chi (S):=\tanh ^{-1}2\sqrt{S(1-S)} \)
and \( \rho (S):=(2S-1)^{-1}|2S-1| \).}
\end{enumerate}
\end{thm}

By the equivalence of norms \( \parallel \cdot \parallel _{S} \) and \( \parallel \cdot \parallel _{S'} \),
we can see a bounded operator \( S' \) on \( K_{S'} \) as a bounded operator
on \( K_{S} \). 

\begin{lm} \emph{If} \( \alpha \parallel \cdot \parallel _{S}\leq \parallel \cdot \parallel _{S'}\leq \beta \parallel \cdot \parallel _{S} \),
\emph{there exists} \( 0<\alpha '<\beta ' \) \emph{such that} \( \alpha '\parallel \cdot \parallel _{P_{S}}\leq \parallel \cdot \parallel _{P_{S'}}\leq \beta '\parallel \cdot \parallel _{P_{S}} \).\end{lm}
\begin{proof} Immediate from \eqref{eq:relation_between_<,>(S)_and_<,>(P_S)}.\end{proof}

\begin{lm}\label{thm:P(S)-P(S'):H.S.<==>1-exp(-x(S))exp(x(S')):H.S.} \emph{Let}
\( S,\, S'\in \mathfrak {S} \) \emph{and} \textit{the topologies induced by
\( \parallel f\parallel _{S} \) and \( \parallel f\parallel _{S'} \) on \( K \)
are equivalent}. \emph{Then the following conditions are equivalent}.

\begin{enumerate}
\item \( P_{S}-P_{S'} \) \emph{is a Hilbert-Schimdt class operator},
\item \( 1-\rho (S)e^{-\chi (S)}e^{\chi (S')}\rho (S') \) \emph{is a Hilbert-Schmidt
class operator},
\item \( 1-\rho (S')e^{-\chi (S')}e^{\chi (S)}\rho (S) \) \emph{is a Hilbert-Schmidt
class operator}.
\end{enumerate}
\end{lm}
\begin{proof} See Lemma 6.5 of \cite{Araki-CCR-2}.\end{proof}

\begin{lm}\label{thm:Unitary_Induced_by_Bogoliubov_trans} \emph{Let} \( S,S'\in \mathfrak {S} \)
\emph{be basis projections and assume that} \( K=K_{S}=K_{S'} \). 

\begin{enumerate}
\item \emph{Let} \( \theta (S,S') \) \emph{be a non-negative hermitian operator on}
\( K \) \emph{satisfying} \( \sinh ^{2}\theta (S,S')=-(S-S')^{2} \). \emph{Let}
\begin{gather*}
   u_{12}(S/S'):=(\sinh \theta (S,S')\cosh \theta (S,S'))^{-1}SS'(1-S), \\
   u_{21}(S/S'):=-(\sinh \theta (S,S')\cosh \theta (S,S'))^{-1}(1-S)S'S, \\
   H(S/S'):=-i\theta (S,S')\{u_{12}(S/S')+u_{21}(S/S')\}. 
\end{gather*}
\emph{Then \( u_{ij}(S/S')^{*}=u_{ji}(S/S') \)} \emph{and} \( H(S/S') \) \emph{satisfies}
\begin{gather*}
   H(S/S')^{\dagger }=H(S/S'),\quad \Gamma H(S/S')\Gamma =-H(S/S'), \\
   (iH(S/S'))^{*}=iH(S/S'). 
\end{gather*}
 ({*} \emph{is relative to \( (\cdot ,\cdot )_{S} \)}.) \emph{Let}
\[
U(S/S'):=\exp (iH(S/S')).
\]
\( U(S/S') \) \emph{satisfies} 
\begin{gather*}
 U(S/S')^{\dagger }U(S/S')=U(S/S')U(S/S')^{\dagger }=1,\quad [\Gamma ,U(S/S')]=0, \\
 U(S/S')^{\dagger }SU(S/S')=S'.  
\end{gather*}
\item \( S-S' \) \emph{is a Hilbert-Schmidt class operator if and only if} \( \theta (S,S') \)
\emph{is a Hilbert-Schmidt class operator}.
\item \emph{Let \( \theta (S,S') \) be a Hilbert-Schmidt class operator. Then there
exists an unique unitary operator} \( T(S,S')\in \mathcal{M}_{S} \) \emph{such
that}
\[
T(S,S')^{*}\overline{\pi _{S}(A)}T(S,S')=\pi _{S}[\tau (U(S/S'))A]\]
\emph{on} \( D(\pi _{S}) \) \emph{and}
\begin{equation}
\label{eq:<w,T(S,S')w>}
\left\langle \Omega _{S},T(S,S')\Omega _{S}\right\rangle =\mathrm{det}_{SK}\left( \frac{1}{\sqrt{\cosh \theta (S,S')}}\right) 
\end{equation}
\emph{where} \( \mathrm{det}_{SK} \) \emph{is the determinant of} \( SK \).(\emph{Since}
\( \theta (S,S') \) \emph{commutes with} \( S \), \emph{the right hand side
of} \eqref{eq:<w,T(S,S')w>} \emph{is well-defined}.)
\end{enumerate}
\end{lm}
\begin{proof} (1) See Lemma 5.4 of \cite{Araki-CCR-2}. (3) See Lemma 5.5 of
\cite{Araki-CCR-2}.\end{proof}

\begin{lm}\label{thm:S-S':H.S-->q(S)_q(S'):unitary_eqiv} \emph{Assume that}
\( S,\, S'\in \mathfrak {S} \) \emph{are basis projections and} \textit{the
topologies induced by \( \parallel f\parallel _{S} \) and \( \parallel f\parallel _{S'} \)
on \( K \) are equivalent}. \emph{If} \( S-S' \) \emph{is a Hilbert-Schmidt
class operator}, \emph{then} \( q_{S}^{\sigma } \) \emph{and} \( q_{S'}^{\sigma } \)
\emph{are unitary equivalent}.\end{lm}
\begin{proof} Let
\[
V\pi _{S'}(A)\Omega _{S'}=\overline{\pi _{S}(A)}T(S,S')\Omega _{S},\quad A\in \mathfrak {A}(K,\gamma ,\Gamma ).\]
Since \( U(S/S')^{\dagger }SU(S/S')=S' \), we have \( \varphi _{S'}=\varphi _{S}\circ \tau (U(S/S')) \).
This shows that \( V \) is an unitary operator from \( \mathcal{H}_{S'} \)
to \( \mathcal{H}_{S} \) and satisfies \( V\pi _{S'}(A)=\overline{\pi _{S}(A)}V \)
on \( D(\pi _{S'}) \) for all \( A\in \mathfrak {A}(K,\gamma ,\Gamma ) \).
By \( V\mathcal{H}_{S'}^{\sigma }\subset \mathcal{H}_{S}^{\sigma } \), the
restriction of \( V \) to \( \mathcal{H}_{S'}^{\sigma } \) is an unitary operator
from \( \mathcal{H}_{S'}^{\sigma } \) to \( \mathcal{H}_{S}^{\sigma } \) and
we have \( VQ^{\sigma }_{S'}(H)=Q_{S}^{\sigma }(H)V \) for all \( H\in sp(\infty ) \).
Thus \( q_{S}^{\sigma } \) is unitary equivalent to \( q_{S'}^{\sigma } \).\end{proof}

The next corollary is directly seen from the above lemma.

\begin{cor}\label{thm:P(S)-S(S'):H.S.=>q_P(S)=q_P(S')} \emph{If} \textit{the
topologies induced by \( \parallel f\parallel _{S} \) and \( \parallel f\parallel _{S'} \)
on \( K \) are equivalent} \emph{and} \( P_{S}-P_{S'} \) \emph{is a Hilbert-Schmidt
class operator}, \emph{then} \( q_{P_{S}}^{\sigma } \) \emph{and} \( q_{P_{S'}}^{\sigma } \)
\emph{are unitary equivalent}.\end{cor} 

\begin{lm} \emph{Assume that} \( S,\, S'\in \mathfrak {S} \) \emph{are not
projection}. \emph{If} \( 1-\rho (S)e^{-\chi (S)}e^{\chi (S')}\rho (S') \)
\emph{is a Hilbert-Schmidt class operator}, \emph{then} \( q_{S}^{\sigma } \)
\emph{and} \( q_{S'}^{\sigma } \) \emph{are quasi-equivalent}. \end{lm}
\begin{proof} Immediate from Corollary \ref{thm:S:non-projection==>q_S=q_P(S)},
Lemma \ref{thm:P(S)-P(S'):H.S.<==>1-exp(-x(S))exp(x(S')):H.S.} and Corollary
\ref{thm:P(S)-S(S'):H.S.=>q_P(S)=q_P(S')}.\end{proof}

\begin{lm} \emph{Let} \( S\in \mathfrak {S} \) \emph{be a projection and} \( S'\in \mathfrak {S} \)
\emph{be not a projection}. \emph{Then} \( 1-\rho (S)e^{-\chi (S)}e^{\chi (S')}\rho (S') \)
\emph{is not a Hilbert-Schmidt class operator}.\end{lm}
\begin{proof} Suppose \( q^{+}_{S}\sim _{q}q^{+}_{S'} \) and \( q_{S}^{-}\sim _{q}q^{-}_{S'} \).
Then since \( q^{+}_{S}\sim _{q}q_{S}^{-} \) and the irreducibility of \( q_{S}^{\sigma } \),
we have \( q^{+}_{S}\sim q_{S}^{-} \). However, since \( S \) is a projection,
\( q_{S}^{+}\not \sim q_{S}^{-} \). This is contradiction. So \( q_{S}^{+}\not \sim _{q}q_{S'}^{+} \)
or \( q_{S}^{-}\not \sim _{q}q_{S'}^{-} \). If \( q_{S}^{+}\sim _{q}q_{S'}^{+} \)
and \( q_{S}^{-}\not \sim _{q}q_{S'}^{-} \), then \( P_{S}-P_{S'} \) is not
a Hilbert-Schmidt class operator from \( q_{P_{S'}}^{+}\sim _{q}q_{S'}^{+}\sim _{q}q_{S}^{+}\not \sim _{q}q^{+}_{P_{S}} \).
Thus \( 1-\rho (S)e^{-\chi (S)}e^{\chi (S')}\rho (S') \) is not a Hilbert-Schmidt
class operator. The case of \( q_{S}^{-}\sim _{q}q^{-}_{S'} \) and \( q_{S}^{+}\not \sim _{q}q_{S'}^{+} \)
is quite similar. If \( q_{S}^{+}\sim _{q}q^{-}_{S'} \) and \( q_{S}^{+}\not \sim _{q}q_{S'}^{+} \)
and \( q_{S}^{-}\not \sim _{q}q_{S'}^{-} \), we have \( q_{P_{S'}}^{+}\sim _{q}q^{-}_{P_{S'}}\sim _{q}q_{S'}^{-}\sim _{q}q_{S}^{+}\not \sim _{q}q_{P_{S}}^{+} \).
Thus \( 1-\rho (S)e^{-\chi (S)}e^{\chi (S')}\rho (S') \) is not a Hilbert-Schmidt
class operator. The case of \( q_{S}^{+}\not \sim _{q}q^{-}_{S'} \) and \( q_{S}^{+}\not \sim _{q}q_{S'}^{+} \)
and \( q_{S}^{-}\not \sim _{q}q_{S'}^{-} \) is trivial. Therefore, \( 1-\rho (S)e^{-\chi (S)}e^{\chi (S')}\rho (S') \)
is not a Hilbert-Schmidt class operator.\end{proof}

From the above lemmas, we have the necessary condition of Theorem \ref{thm:Main_Theorem}.

\begin{lm}[Necessity of Theorem \ref{thm:Main_Theorem}] \emph{Suppose that}
\( S,\, S'\in \mathfrak {S} \) \emph{satisfy the following two conditions}.

\begin{enumerate}
\item \textit{The topologies induced by \( \parallel f\parallel _{S} \) and \( \parallel f\parallel _{S'} \)
on \( K \) are equivalent},
\item \textit{\( 1-\rho (S)e^{-\chi (S)}e^{\chi (S')}\rho (S') \) is a Hilbert-Schmidt
class operator.}
\end{enumerate}
\emph{Then two representations} \( q_{S}^{\sigma } \) \emph{and} \( q_{S'}^{\sigma } \)
\emph{are quasi-equivalent}. \end{lm}

Next, we prove the sufficiency of Theorem \ref{thm:Main_Theorem}.

A state \( \varphi _{P_{S}} \) on CCR algebra \( \mathfrak {A}(K_{S}\oplus K_{S},\widehat{\gamma }_{S},\widehat{\Gamma }_{S}) \)
can be viewed as a state on \( \mathcal{M}_{S}^{+} \) satisfying \( \varphi _{P_{S}}(Q):=\left\langle \Omega _{P_{S}},Q\Omega _{P_{S}}\right\rangle  \),
\( Q\in \mathcal{M}_{P_{S}}^{+} \). 

Now let \( \dim K<\infty  \). Since \( P_{S}-P_{S'} \) is a Hilbert-Schmidt
class operator, \( q_{P_{S}}^{+} \) and \( q_{P_{S'}}^{+} \) are unitary equivalent
and we can identify \( \mathcal{M}_{P_{S}}^{+} \) with \( \mathcal{M}_{P_{S'}}^{+} \).
Therefore, a state \( \varphi _{P_{S'}} \) on CCR algebra \( \mathfrak {A}(K_{S'}\oplus K_{S'},\widehat{\gamma }_{S'},\widehat{\Gamma }_{S'}) \)
is regarded as a state on \( \mathcal{M}^{+}_{P_{S}} \) satisfying \( \varphi _{P_{S'}}(Q):=\left\langle \Omega ',Q\Omega '\right\rangle  \),
\( \Omega ':=T(P_{S},P_{S'})\Omega _{P_{S}} \), \( Q\in \mathcal{M}^{+}_{P_{S'}}=\mathcal{M}^{+}_{P_{S}} \)
where \( T(P_{S},P_{S'}) \) is an unitary operator determined by (3) of Lemma
\ref{thm:Unitary_Induced_by_Bogoliubov_trans}.

We quote the following two results to prove the sufficiency of the main theorem.

\begin{lm}\label{thm:THEOREM4(5)} \emph{Let} \( \mathcal{M} \) \emph{be a
von Neumann algebra on a Hilbert space \( \mathcal{H} \) with a cyclic and
separating vector} \( \Psi  \). \emph{Let} \( V_{\Psi } \) \emph{be a natural
positive corn associated with the pair} \( (\mathcal{M},\Psi ) \). \emph{If}
\( \Phi  \) \emph{is an another cyclic and separating vector for} \( \mathcal{M} \),
\emph{then} \( \Phi \in V_{\Psi } \) \emph{if and only if the following 2 conditions
hold} : 
\begin{enumerate}
\item \( J_{\Phi }=J_{\Psi } \),
\item \( \left\langle \Phi ,Q_{+}\Psi \right\rangle \geq 0 \) \emph{for all} \( Q_{+}\in \mathcal{M}\cap \mathcal{M}',\, Q_{+}\geq 0 \).
\end{enumerate}
\end{lm}
\begin{proof} See THEOREM 4 (5) of \cite{Radon-Nikodym-theorem}.\end{proof}

\begin{lm}\label{thm:THEOREM4(8)} \emph{Let} \( \mathcal{M} \) \emph{be a
von Neumann algebra on a Hilbert space \( \mathcal{H} \) with a cyclic and
separating vector} \( \Psi  \) \emph{and let \( V_{\Psi } \) be the natural
positive corn for \( \Psi  \). Let \( \Phi _{i}\in V_{\Psi }(i=1,2) \) and
\( \varphi _{\Phi _{i}} \) be a vector state for \( \Phi _{i} \). Then}
\[
\parallel \varphi _{\Phi _{1}}-\varphi _{\Phi _{2}}\parallel \geq \parallel \Phi _{1}-\Phi _{2}\parallel ^{2}.\]
\end{lm}
\begin{proof} See THEOREM 4 (8) of \cite{Radon-Nikodym-theorem}.\end{proof}

\begin{lm} \emph{Let \( \dim K<\infty  \) and \( S,\, S'\in \mathfrak {S} \)
be} \( 0<S<1 \), \( 0<S'<1 \). \emph{Then}
\begin{equation}
\label{eq:||s(P_S)-s(P_S')||>||w-w'||^2}
\parallel (\varphi _{P_{S}}-\varphi _{P_{S'}})|\mathcal{M}^{+}_{P_{S}}\parallel \geq 2\left\{ 1-\mathrm{det}_{P_{S}K_{P_{S}}}\left( \frac{1}{^{4}\sqrt{P_{S}P_{S'}P_{S}}}\right) \right\} 
\end{equation}
\end{lm}
\begin{proof} Let \( J_{\Omega '} \) and \( \Delta _{\Omega '} \) be the modular
conjugation and modular operator associated with 
the pair \( (\mathcal{M}^{+}_{P_{S}},\Omega ') \)
and let \( V_{\Omega _{P_{S}}} \) be the natural positive corn associated with
the pair \( (\mathcal{M}_{P_{S}}^{+},\Omega _{P_{S}}) \). 
We show \( J_{\Omega '}=J_{\Omega _{P_{S}}} \)
and \( \left\langle \Omega ',Q\Omega _{P_{S}}\right\rangle \geq 0 \) for all
\( Q\in C(\mathcal{M}_{P_{S}}^{+}),\, Q\geq 0 \) with help of Lemma \ref{thm:THEOREM4(5)}
to prove \( \Omega '\in V_{\Omega _{P_{S}}} \) . 

We prove the first part, \( J_{\Omega '}=J_{\Omega _{P_{S}}} \). 
Since we have
\[
  [J_{\Omega _{P_{S}}},T(P_{S},P_{S'})]=0
\]
(See (6.2) of \cite{Araki-CCR-2}), the following relation holds :
\begin{equation}
J_{\Omega _{P_{S}}}\Omega '=\Omega '. \label{eq:T(P,P')x=x}
\end{equation}
We remark that we have already obtained the following relations :
\begin{gather}
\label{eq:VQV*=Q'}
  V^* Q_{P_S}(H)V=Q_{P_{S'}}(H),\\
\label{eq:V*JV=J'}
  J_{\Omega '} =V J_{\Omega _{P_{S'}}} V^*, \\
\label{eq:JQJ=Q'(Omega)}
  J_{\Omega _{P_S}} Q_{P_S}(H) J_{\Omega _{P_S}}={\widehat{Q}}_{P_S}(0\oplus H)^{*},\\
\label{eq:JQ(H)J=Q(0+H)(P_S')}
  J_{\Omega _{P_{S'}}} Q_{P_{S'}}(H) J_{\Omega _{P_{S'}}}=\widehat{Q}_{P_{S'}}(0\oplus H)^*
\end{gather}
where \( V\) is the unitary operator defined in Lemma
\ref{thm:S-S':H.S-->q(S)_q(S'):unitary_eqiv}. We have
\begin{equation}
  J_{\Omega '} Q_{P_S} (H) J_{\Omega '}=\widehat{Q} _{P_S}(0\oplus H)^*
  \label{eq:JQJ=Q'(Omega')}
\end{equation}
from \eqref{eq:VQV*=Q'}, \eqref{eq:V*JV=J'} and \eqref{eq:JQ(H)J=Q(0+H)(P_S')}. 
It follows
\begin{gather*}
[J_{\Omega '}J_{\Omega _{P_S}}, Q_{P_S}(H)]=0,\\
[J_{\Omega '}J_{\Omega _{P_S}}, \widehat{Q}_{P_S}(0\oplus H)]=0
\end{gather*}
for all \( H\in sp(\infty )\) 
from \eqref{eq:JQJ=Q'(Omega)} and \eqref{eq:JQJ=Q'(Omega')}.
Now the center \( C(\mathcal{M}_{P_S} )\) of \( \mathcal{M}_{P_S}\) is trivial :
\[
C(\mathcal{M}_{P_S})=
\left(
  \begin{array}{cc}
    \mathcal{M}^+_{P_S} &         0            \\
            0         &    \mathcal{M}^-_{P_S}
  \end{array}
\right)
\cap \mathcal{M}'_{P_S}= 
\left(
  \begin{array}{cc}
    C( \mathcal{M}^+_{P_S} ) &       0              \\
            0         &    C(\mathcal{M}^-_{P_S} )
  \end{array}
\right)
=\mathbf{C} 1.
\]
Thus \( J_{\Omega '}J_{\Omega _{P_S}} \in
C(\mathcal{M}_{P_S} )=\mathbf{C} 1\), that is,
\begin{equation}
J_{\Omega '}=\lambda J_{\Omega _{P_S}},\quad \lambda \in 
S^1:=\{ \lambda \in \mathbf{C} \ |\ |\lambda |=1\}.
\label{eq:J=J'}
\end{equation}
Due to \eqref{eq:T(P,P')x=x} and \eqref{eq:J=J'},  \( \lambda =1\).

The second part is verified by \eqref{eq:<w,T(S,S')w>} and the factoriality of
\( \mathcal{M}_{P_{S}}^{+} \) and 
\[
\left\langle \Omega ',\Omega _{P_{S}}\right\rangle =\left\langle \Omega _{P_{S}},T(P_{S},P_{S'})\Omega _{P_{S}}\right\rangle =\prod _{\lambda \in \sigma (\theta (P_{S},P_{S'}))}\frac{1}{\sqrt{\cosh \lambda }}\geq 0.\]
 Therefore \( \Omega '\in V_{\Omega _{P_{S}}} \). Now from \eqref{eq:<w,T(S,S')w>}
and Lemma \ref{thm:THEOREM4(8)} and
\[\begin{split}
P_{S}\cosh \theta (P_{S},P_{S'}) & = P_{S}\sqrt{1-[\sinh \theta (P_{S},P_{S'})]^{2}}\\
 & = \sqrt{P_{S}\{1-(P_{S}-P_{S'})^{2}\}}\\
 & = \sqrt{P_{S}P_{S'}P_{S}},
\end{split}\]
we obtain \eqref{eq:||s(P_S)-s(P_S')||>||w-w'||^2}.\end{proof}

\begin{lm}\label{thm:lim||(s-s')|M+||=2} \emph{Assume that \( K \) is separable.
Let} \( S,\, S'\in \mathfrak {S} \) \emph{be} \( 0<S<1 \), \( 0<S'<1 \) \emph{an}\textit{d}
\emph{}\textit{the topologies induced by \( \parallel f\parallel _{S} \) and
\( \parallel f\parallel _{S'} \) on \( K \) are equivalent}. \emph{If} \( P_{S}-P_{S'} \)
\emph{is not a Hilbert-Schmidt class operator}, \emph{then there exists \( \Gamma  \)-invariant
finite dimensional subspaces \( K_{n} \) of \( K \) such that} 
\[
\lim _{n\rightarrow \infty }\parallel (\varphi _{S_{n}}-\varphi _{S'_{n}})|\mathcal{M}^{+}_{P_{S_{n}}}\parallel =2.\]
\emph{\( S_{n}(f,g) \) is the restriction of \( S(f,g) \) to \( K_{n} \)}.\end{lm}
\begin{proof} Since we have the inequality \eqref{eq:||s(P_S)-s(P_S')||>||w-w'||^2},
we can verify this lemma as the proof of Lemma 6.7 of \cite{Araki-CCR-2}. \end{proof}

\begin{lm}[Sufficiency of Theorem \ref{thm:Main_Theorem}]\label{thm:Sufficiency_of_Main_Theorem}
\emph{Assume that \( K \) is separable}. \( S,\, S'\in \mathfrak {S} \) \textit{and
the topologies induced by \( \parallel f\parallel _{S} \) and \( \parallel f\parallel _{S'} \)
on \( K \) are equivalent}. \emph{If \( 1-\rho (S)e^{-\chi (S)}e^{\chi (S')}\rho (S') \)
is not a Hilbert-Schmidt class operator}, \emph{then \( q_{S}^{\sigma } \)
and \( q_{S'}^{\sigma } \) are not quasi-equivalent}.\end{lm}
\begin{proof} First suppose that \( 0<S<1 \) and \( 0<S'<1 \). Then we can
show \( q_{S}^{+}\not \sim _{q}q_{S'}^{+} \) from Lemma \ref{thm:lim||(s-s')|M+||=2}
as the proof of Lemma 6.8 of \cite{Araki-CCR-2}. Since \( q_{S}^{-}\sim _{q}q_{S}^{+} \)
and \( q_{S'}^{-}\sim _{q}q_{S'}^{+} \) by Lemma 6.2, we have \( q_{S}^{-}\not \sim _{q}q_{S'}^{-} \)
as well. 

For general \( S \) and \( S' \), this proof is quite similar to Lemma 6.8
of \cite{Araki-CCR-2}.\end{proof}

\section{Metaplectic Representations of \protect\( Sp(\infty ,P)\protect \)}

\begin{defini} \emph{A bijective linear map} \( U \) \emph{on} \( K \) \emph{satisfies}
\( \gamma (Uf,Ug)=\gamma (f,g) \) \emph{and} \( \Gamma U=U\Gamma  \). \emph{Let}
\( \tau (U) \) \emph{be a} {*}-\emph{automorphism of CCR algebra} \( \mathfrak {A}(K,\gamma ,\Gamma ) \)
\emph{satisfying} \( \tau (U)B(f)=B(Uf) \). \emph{Then we call} \( U \) \emph{a
Bogoliubov transformation for \( (K,\gamma ,\Gamma ) \) and} \emph{\noun{\( \tau (U) \)}}
\emph{a Bogoliubov} {*}-\emph{automorphism}. \emph{The set of all Bogoliubov
transformations is called the} \emph{symplectic group} \emph{and we denote the
symplectic group for} \( (K,\gamma ,\Gamma ) \) \emph{by} \( Sp(K,\gamma ,\Gamma ) \).
\end{defini}

We define some subgroups of \( Sp(K,\gamma ,\Gamma ) \). 

Let \( Sp(\infty ) \) be the group generated by \( sp(\infty ) \) :
\[
Sp(\infty ):=\{e^{iH}\, |\, H\in sp(\infty )\}.\]

Suppose that \( P \) is a basis projection for \( (K,\gamma ,\Gamma ) \) satisfying
\( K_{P}=K \). 

Let \( Sp(\infty ,P) \) be the set of all \( U\in Sp(K,\gamma ,\Gamma ) \)
satisfying
\begin{equation}
\label{eq:boundedness_HS_norm_of_PU(1-P)}
\left\Vert PU(1-P)\right\Vert _{\mathrm{H}.\mathrm{S}.}<\infty 
\end{equation}
where \( \left\Vert \cdot \right\Vert _{\mathrm{H}.\mathrm{S}.} \) is the Hilbert-Schmidt
norm of operators on \( K \). \eqref{eq:boundedness_HS_norm_of_PU(1-P)} is the
sufficient condition of the existence of unitary representation of \( U \).

\begin{lm}\label{thm:property_of_Sp(n)} \emph{Let}
\[
d_{P}(A,B):=\left\Vert A-B\right\Vert +\left\Vert P(A-B)(1-P)\right\Vert _{\mathrm{H}.\mathrm{S}.}\]
\emph{for all} \( A,B\in Sp(\infty ,P) \). \emph{Then} \( Sp(\infty ,P) \)
\emph{is a topological group with respect to} \( d_{P} \). 

\end{lm}

\begin{proof} We have to show the continuity of multiplication and inverse.
Suppose that \( A,A_{\nu },B,B_{\nu }\in Sp(\infty ,P) \) satisfy
\begin{equation}
\label{eq:d(A_n,A)->0}
d_{P}(A_{\nu },A)\rightarrow 0,\quad d_{P}(B_{\nu },B)\rightarrow 0.
\end{equation}
\eqref{eq:d(A_n,A)->0} says that for any \( \varepsilon >0 \), if \( \nu \rightarrow \infty  \),
then 
\begin{gather}
 \left\Vert PB_{\nu }(1-P)\right\Vert _{\mathrm{H}.\mathrm{S}.}\leq \left\Vert PB(1-P)\right\Vert _{\mathrm{H}.\mathrm{S}.}+\varepsilon, \label{eq:boundedness_of_net_PB(1-P)} \\
 \left\Vert B_{\nu }\right\Vert \leq \left\Vert B\right\Vert +\varepsilon . \label{eq:boundedness_of_net_B} 
\end{gather}

First we claim
\begin{equation}
\label{eq:convergence_of_PAB(1-P)}
\left\Vert P(A_{\nu }-A)B(1-P)\right\Vert _{\mathrm{H}.\mathrm{S}.}\rightarrow 0.
\end{equation}
\eqref{eq:convergence_of_PAB(1-P)} follows from \eqref{eq:boundedness_of_net_PB(1-P)}
and
\[\begin{split}
\| P(&A_{\nu} -A)B(1-P) \|_{\mathrm{H.S.}}\\
&\le \| P(A_{\nu}-A)(1-P)B(1-P)\| _{\mathrm{H.S.}}+
\| P(A_{\nu} -A)PB(1-P)\| _{\mathrm{H.S.}}\\
&\le \| P(A_{\nu}-A)(1-P)\| _{\mathrm{H.S.}}\|B\| +
\|A_{\nu}-A\| \|PB(1-P)\|_{\mathrm{H.S.}} .
\end{split}\]

Due to \eqref{eq:convergence_of_PAB(1-P)}, we have \( d_{P}(A_{\nu }B_{\nu },AB)\rightarrow 0 \)
as \( \nu \rightarrow \infty  \).

On the other hand, Since \eqref{eq:boundedness_of_net_PB(1-P)}, \eqref{eq:boundedness_of_net_B}
and
\[\begin{split}
\| P(&A_{\nu}^{-1} -A^{-1} (1-P)\| _{\mathrm{H.S.}}\\
&=\| PA_{\nu}^{-1} (A_{\nu}-A)A^{-1}(1-P)\|_{\mathrm{H.S.}}\\
&\le \| PA_{\nu}^{-1} (1-P)(A_{\nu}-A)A^{-1}(1-P)\|_{\mathrm{H.S.}}
+\| PA_{\nu}^{-1} P(A_{\nu}-A)A^{-1}(1-P)\|_{\mathrm{H.S.}}\\
&\le \| PA_{\nu}^{-1} (1-P)\|_{\mathrm{H.S.}} \| A_{\nu}-A\| \| A^{-1}\|+
\|A_{\nu}^{-1}\| \|P(A_{\nu}-A)A^{-1}(1-P)\|_{\mathrm{H.S.}},
\end{split}\] 
we have \( d_{P}(A_{\nu }^{-1},A^{-1})\rightarrow 0 \) as \( \nu \rightarrow \infty  \).
\end{proof}

\begin{lm} \emph{Assume that} \( P_{1} \) \emph{and} \( P_{2} \) \emph{are
basis projections and there exist} \( \beta >\alpha >0 \) \emph{such that}
\( \alpha \left\Vert f\right\Vert _{P_{1}}\leq \left\Vert f\right\Vert _{P_{2}}\leq \beta \left\Vert f\right\Vert _{P_{1}} \)
\emph{for all} \( f\in K \) \emph{and} \( P_{1}-P_{2} \) \emph{is a Hilbert-Schmidt
class operator}. \emph{Then} \( Sp(\infty ,P_{1})=Sp(\infty ,P_{2}) \) \emph{as
a topological group}. \end{lm}

\begin{proof} We have only to show the equivalence of the distance \( d_{P_{1}} \)
and \( d_{P_{2}} \). In this proof, we denote the operator norm with respect
to \( P \) by \( \left\Vert \cdot \right\Vert _{P} \) and the Hilbert-Schmidt
norm with respect to \( P \) by \( \left\Vert \cdot \right\Vert _{\mathrm{H}.\mathrm{S}.,P} \).
(In this proof, the Hilbert space norm of \( K \) and the operator norm of
bounded operators on \( K \) is same notation, however we probably does not
confuse the two meanings.)
\[
\left\Vert A\right\Vert _{P_{2}}:=\sup _{x\in K}\frac{\left\Vert Ax\right\Vert _{P_{2}}}{\left\Vert x\right\Vert _{P_{2}}}\leq \sup _{x\in K}\frac{\beta \left\Vert Ax\right\Vert _{P_{1}}}{\alpha \left\Vert x\right\Vert _{P_{1}}}=\frac{\beta }{\alpha }\left\Vert A\right\Vert _{P_{1}}.\]
In the same way, we have \( \left\Vert A\right\Vert _{P_{1}}\leq \frac{\beta }{\alpha }\left\Vert A\right\Vert _{P_{2}} \).

On the other hand, 
\[\begin{split}
\| P_2 &A(1-P_2)\| _{\mathrm{H.S.},P_2} \\ 
&\le \beta \| P_2U(P_1/P_2)^{\dagger} U(P_1/P_2)A
U(P_1/P_2)^{\dagger} U(P_1/P_2)(1-P_2)\| _{\mathrm{H.S.},P_1} \\
&\le \beta \| U(P_1/P_2)^{\dagger}P_1U(P_1/P_2)A
U(P_1/P_2)^{\dagger}(1-P_1)U(P_1/P_2)\| _{\mathrm{H.S.},P_1}\\
&\le \beta \| U(P_1/P_2)^{\dagger}\|_{P_1} \| U(P_1/P_2)\|_{P_1}\\ 
&\qquad \times \| P_1U(P_1/P_2)A
U(P_1/P_2)^{\dagger}(1-P_1)\| _{\mathrm{H.S.},P_1}\end{split}\] 
and
\[\begin{split}
\| P_1&U(P_1/P_2)AU(P_1/P_2)^{\dagger}(1-P_1)\| _{\mathrm{H.S.},P_1}\\
&\le \| P_1U(P_1/P_2)\{ P_1+(1-P_1)\} A
U(P_1/P_2)^{\dagger}(1-P_1)\| _{\mathrm{H.S.},P_1}\\
&\le \| U(P_1/P_2)\| _{P_1}
\| P_1AU(P_1/P_2)^{\dagger}(1-P_1)\| _{\mathrm{H.S.},P_1}\\
&\qquad +\| P_1U(P_1/P_2)(1-P_1)\| _{\mathrm{H.S.},P_1}
\| U(P_1/P_2)^{\dagger}\| _{P_1}\| A\|_{P_1}\\
&\le \| U(P_1/P_2)\| _{P_1}\{ \| A\| _{P_1}
\| P_1U(P_1/P_2)^{\dagger}(1-P_1)\| _{\mathrm{H.S.},P_1}\\ 
&\qquad + \| P_1A(1-P_1)\| _{\mathrm{H.S.},P_1}
\| U(P_1/P_2)^{\dagger}\| _{P_1} \}\\
&\qquad +\| P_1U(P_1/P_2)(1-P_1)\| _{\mathrm{H.S.},P_1}
\| U(P_1/P_2)^{\dagger}\| _{P_1}\| A\|_{P_1}\\ 
&\le M''(\| A\|_{P_1}+\| P_1A(1-P_1)\| _{\mathrm{H.S.},P_1})
\end{split}\]
where \( M'' \) is the maximum value of
\begin{gather*}
 \left\Vert P_{1}U(P_{1}/P_{2})(1-P_{1})\right\Vert _{\mathrm{H}.\mathrm{S}.,P_{1}}\left\Vert U(P_{1}/P_{2})^{\dagger }\right\Vert _{P_{1}}, \\
 \left\Vert P_{1}U(P_{1}/P_{2})^{\dagger }(1-P_{1})\right\Vert _{\mathrm{H}.\mathrm{S}.,P_{1}}\left\Vert U(P_{1}/P_{2})\right\Vert _{P_{1}}, \\
 \left\Vert U(P_{1}/P_{2})^{\dagger }\right\Vert _{P_{1}}\left\Vert U(P_{1}/P_{2})\right\Vert _{P_{1}}.
\end{gather*}
Since \( P_{1}U(P_{1}/P_{2})(1-P_{1})=\sinh \theta (P_{1},P_{2})u_{12}(P_{1}/P_{2}) \),
\( P_{1}U(P_{1}/P_{2})(1-P_{1}) \) is a Hilbert-Schmidt class operator and
\( M'' \) is not infinity. From the above argument,
\[
\left\Vert P_{2}A(1-P_{2})\right\Vert _{\mathrm{H}.\mathrm{S}.,P_{2}}\leq M'(\left\Vert A\right\Vert _{P_{1}}+\left\Vert P_{1}A(1-P_{1})\right\Vert _{\mathrm{H}.\mathrm{S}.,P_{1}})\]
where \( M':=\beta \left\Vert U(P_{1}/P_{2})^{\dagger }\right\Vert _{P_{1}}\left\Vert U(P_{1}/P_{2})\right\Vert _{P_{1}}M'' \).

Let \( M:=M'+\beta \alpha ^{-1} \). Then \( d_{P_{2}}(U_{1},U_{2})\leq Md_{P_{1}}(U_{1},U_{2}) \).
In the same way, we can show that there exists a positive number \( m>0 \)
such that \( d_{P_{1}}(U_{1},U_{2})\leq \frac{1}{m}d_{P_{2}}(U_{1},U_{2}) \).
Therefore
\[
md_{P_{1}}(U_{1},U_{2})\leq d_{P_{2}}(U_{1},U_{2})\leq Md_{P_{1}}(U_{1},U_{2}).\]
\end{proof}

\begin{lm} \emph{Let} \( U\in Sp(\infty ,P) \) \emph{and} \( P':=UPU^{\dagger } \).

\begin{enumerate}
\item \emph{Let} \( R(U):=U(P/P')U \). \emph{Then} \( R(U) \) \emph{commutes with}
\( P \) \emph{and} \( R(U) \) \emph{is an unitary operator on} \( K \).
\item \( U(P/P')^{\dagger } \) \emph{is a positive and} 1 + \emph{Hilbert-Schmidt
class operator}.
\end{enumerate}
\end{lm}

\begin{proof} (1) Since \( P'=UPU^{\dagger } \) and \( PU(P/P')=U(P/P')P' \),
we have
\[
PR(U)=U(P/P')P'U=U(P/P')UPU^{\dagger }U=R(U)P.\]
Thus \( R(U) \) commutes with \( P \).
Due to \( \gamma (R(U)f,R(U)g)=\gamma (f,g) \), we have
\[
\gamma _{P}=R(U)^{*}\gamma _{P}R(U)=\gamma _{P}R(U)^{*}R(U).\]
({*} is relative to \( (\cdot ,\cdot )_{P} \).) Since \( P \) is a projection
and \( \gamma _{P}^{2}=1 \), we obtain
\[
1=\gamma _{P}^{2}=\gamma _{P}\cdot \gamma _{P}R(U)^{*}R(U)=R(U)^{*}R(U).\]
This implies \( R(U)R(U)^{*}=1 \). Thus \( R(U) \) is an unitary operator
on \( K \). 

(2) Due to \eqref{eq:boundedness_HS_norm_of_PU(1-P)}, \( \theta (P,P') \) defined
in Lemma \ref{thm:Unitary_Induced_by_Bogoliubov_trans} is a Hilbert-Schmidt
class operator. Indeed, we have
\begin{equation}
\label{eq:H.S.norm_PU(1-P)_equal_tr(PU(1-P)U*P)}
\left\Vert PU(1-P)\right\Vert ^{2}_{\mathrm{H}.\mathrm{S}.}=\mathrm{tr}(PU(1-P)U^{\dagger }P)
\end{equation}
and
\begin{gather}
 PU(1-P)U^{\dagger }P=-[\sinh \theta (P,P')]^{2}P, \label{eq:PU(1-P)U*P} \\
 \Gamma \cdot PU(1-P)U^{\dagger }P\cdot \Gamma =-[\sinh \theta (P,P')]^{2}(1-P) \label{eq:PU(1-P)U*P_Gamma} 
\end{gather}
from the direct computation. (\eqref{eq:PU(1-P)U*P_Gamma} follows from \( [\sinh \theta (P,P'),\Gamma ]=0. \))
\eqref{eq:H.S.norm_PU(1-P)_equal_tr(PU(1-P)U*P)},
\eqref{eq:PU(1-P)U*P} and  \eqref{eq:PU(1-P)U*P_Gamma}
say that \( \sinh \theta (P,P') \) is a Hilbert-Schmidt class operator, i.e.
\( \theta (P,P') \) is a Hilbert-Schmidt class operator. We obtain immediately
that \( H(P/P') \) is a Hilbert-Schmidt class operator.

Since \( iH(P/P') \) is a hermitian operator, the positivity of \( U(P/P') \)
is obvious.\end{proof}

From the above lemma, \( U\in Sp(\infty ,P) \) is written as \( U=U(P/P')^{\dagger }R(U) \)
and this is the polar decomposition of \( U \).

We introduce some notations to define the metaplectic representations of \( Sp(\infty ,P) \).

Let \( P \) be a basis projection and \( U \) be the element of \( Sp(K,\gamma ,\Gamma ) \)
satisfying \( [P,U]=0 \). (Since \( [P,U]=0 \), \( U \) is an unitary operator.)
Then the operator \( T_{P}(U) \) on \( \mathcal{H}_{P} \) is defined by 
\[
T_{P}(U)\pi _{P}(A)\Omega _{P}:=\pi _{P}(\tau (U)A)\Omega _{P},\quad A\in \mathfrak {A}(K,\gamma ,\Gamma ).\]
 \( T_{P}(U) \) is the second quantization of \( U \). Since \( \tau (U) \)
is a {*}-automorphism of CCR algebra \( \mathfrak {A}(K,\gamma ,\Gamma ) \)
and \( \varphi _{P} \) is a quasifree state satisfying
\[
\varphi _{P}(\tau (U)[B(f)^{*}B(g)])=(Uf,PUg)_{P}=(f,Pg)_{P}=\varphi _{P}(B(f)^{*}B(g)),\]
\( T_{P}(U) \) is an unitary operator on \( \mathcal{H}_{P} \).

Let \( T_{P}(\Gamma ) \) be an antiunitary operator on \( \mathcal{H}_{P} \)
defined by
\[
T_{P}(\Gamma )\pi _{P}(A)\Omega _{P}=\pi _{P}(\tau (\Gamma )A)\Omega _{P},\quad A\in \mathfrak {A}(K,\gamma ,\Gamma ).\]

\begin{lm} \emph{Let} \( U\in Sp(\infty ,P) \). \emph{Then the unitary operator}
\( Q_{P}(U) \) \emph{satisfying}
\begin{equation}
\label{eq:Bogoliubov_trans_implement_unitary}
Q_{P}(U)W_{P}(f)Q_{P}(U)^{*}=W_{P}(Uf)
\end{equation}
\emph{for all} \( f\in \mathrm{Re}K \) \emph{exists uniquely up to} \( S^{1}:=\{\lambda \in \mathbf{C}\, |\, |\lambda |=1\} \). \end{lm}

\begin{proof} Let 
\[
Q_{P}(U):=T(P,P')T_{P}(R(U)).\]
Then \( Q_{P}(U) \) satisfies \eqref{eq:Bogoliubov_trans_implement_unitary}. 

The uniqueness of \( Q_{P}(U) \) follows from the irreducibility of the von
Neumann algebra \( \mathcal{R}_{P}(\mathrm{Re}K) \). In fact, if \( Q'_{P}(U) \)
is an another unitary operator satisfying \eqref{eq:Bogoliubov_trans_implement_unitary},
then we have
\[
Q'_{P}(U)^{*}Q_{P}(U)W_{P}(f)Q_{P}(U)^{*}Q'_{P}(U)=W_{P}(f)\]
for all \( f\in \mathrm{Re}K \) and this shows
\[
Q'_{P}(U)^{*}Q_{P}(U)\in \mathcal{R}_{P}(\mathrm{Re}K)'=\mathbf{C}1.\]
Therefore \( Q_{P}(U) \) is unique up to the phase factor. \end{proof}

\emph{Remark}. Since \( Q_{P}(H) \) defined in the section 4 satisfies \eqref{eq:Q(H)W(f)Q(H)*--projective_case}
and \( Q_{P}(U) \) is unique up to the phase factor, we have \( Q_{P}(e^{iH})=\lambda Q_{P}(H) \),
\( \lambda \in S^{1} \). 

Let
\[
Q_{P}(\lambda ,U):=\lambda Q_{P}(U)\]
for all \( U\in Sp(\infty ,P) \) and \( \lambda \in S^{1} \).

\begin{defini} \emph{We denote the group generated by all} \( Q_{P}(\lambda ,U) \)
\emph{satisfying}
\begin{equation}
\label{eq:T(Gamma)_commute_Q_P(lambda,U)}
[T_{P}(\Gamma ),Q_{P}(\lambda ,U)]=0
\end{equation}
\emph{by} \( Mp(\infty ,P) \). \emph{We call} \( Mp(\infty ,P) \) \emph{the
metaplectic group of} \( Sp(\infty ,P) \). \end{defini}

The elements \( \lambda  \) in \( S^{1} \) satisfying \eqref{eq:T(Gamma)_commute_Q_P(lambda,U)}
are \( 1 \) and \( -1 \). In fact, by \( [\Gamma ,U(P/P')]=0 \) and \( [\Gamma ,R(U)]=0 \),
we have \( [T_{P}(\Gamma ),T(P,P')]=0 \) and \( [T_{P}(\Gamma ),T_{P}(R(U))]=0 \).
This shows \( [T_{P}(\Gamma ),Q_{P}(1,U)]=0 \). Thus
\[
T_{P}(\Gamma )Q_{P}(\lambda ,U)=\overline{\lambda }T_{P}(\Gamma )Q_{P}(1,U)=\overline{\lambda }\lambda ^{-1}Q_{P}(\lambda ,U)T_{P}(\Gamma ).\]
Due to \eqref{eq:T(Gamma)_commute_Q_P(lambda,U)}, \( \overline{\lambda }\lambda ^{-1}=1 \).
Therefore \( \lambda \in S^{1}\cap \mathbf{R}=\{\pm 1\}. \)

\begin{prop}\label{thm:property_of_metaplectic_rep}

\begin{enumerate}
\item \emph{The metaplectic representation \( Mp(\infty ,P) \) is a topological group
with respect to the strong operator topology}.
\item \emph{The metaplectic representation is continuous projective representation
with respect to the topology induced by the distance \( d_{P} \) and the strong
operator topology}, \emph{i}.\emph{e}. \emph{if} \( d_{P}(U_{\nu },U)\rightarrow 0 \)
\emph{as} \( \nu \rightarrow \infty  \), \emph{then} \( Q_{P}(\lambda ,U_{\nu })\rightarrow Q_{P}(\lambda ,U) \)
\emph{strongly}.
\item \( Mp(\infty ,P) \) \emph{is double covering of} \( Sp(\infty ,P). \)
\end{enumerate}
\end{prop}

\begin{proof} (1) This claim is easily checked. 

(2) We prove that \( \left\Vert \{Q_{P}(1,U)-1\}\Omega _{P}\right\Vert ^{2}\rightarrow 0 \)
if \( d_{P}(U,1)\rightarrow 0 \). Since
\[\begin{split}
\mathrm{det}_{P}(\cosh \theta (P,P'))^{2} & \leq  \exp \left( \left\Vert (\cosh \theta (P,P'))^{2}\right\Vert _{\mathrm{tr}}\right) \\
 & = \exp \left( \left\Vert -P(P-P')^{2}P\right\Vert _{\mathrm{tr}}\right) \\
 & = \exp \left( \left\Vert PU(1-P)U^{\dagger }P\right\Vert _{\mathrm{tr}}\right) \\
 & = \exp \left( \left\Vert PU(1-P)\right\Vert _{\mathrm{H}.\mathrm{S}.}\right) ,
\end{split}\]
we have
\[\begin{split}
\left\Vert \{Q_{P}(1,U)-1\}\Omega _{P}\right\Vert ^{2} & = 2(1-\mathrm{Re}\left\langle \Omega _{P},T(P,P')\Omega _{P}\right\rangle )\\
 & = 2\left\{ 1-\mathrm{det}_{PK}\left( \frac{1}{\sqrt{\cosh \theta (P,P')}}\right) \right\} \\
 & = 2\left\{ 1-\frac{1}{^{4}\sqrt{\mathrm{det}_{P}(\cosh \theta (P,P'))^{2}}}\right\} \\
 & \leq  2\left\{ 1-\exp \left( -\frac{1}{4}\left\Vert PU(1-P)\right\Vert _{\mathrm{H}.\mathrm{S}.}\right) \right\} .
\end{split}\]
Thus the first claim has been proved.

Moreover, for any \( f\in \mathrm{Re}K \),
\[\begin{split}
\| \{&Q_P (1,U)-1\} W_P(f) \Omega _P \| \\
&=\| \{ W_P(U^{\dagger} f)Q_P(1,U)-W_P(f)\} \Omega_P\| \\
&\le \| \{ W_P(U^{\dagger} f)Q_P(1,U) -W_P(U^{\dagger}f)\} \Omega _P\| +
\| \{W_P(U^{\dagger}f)-W_P(f)\}\Omega_P\|\\
&\le \| \{Q_P(1,U)-1\} \Omega_P \| +\| \{W_P(U^{\dagger}f)-W_P(f)\}\Omega_P\|
\end{split}\]

From Lemma \ref{thm:Fock-rep-property}(3), if \( \left\Vert U-1\right\Vert \rightarrow 0 \),
we have \( \left\Vert \{W_{P}(U^{\dagger }f)-W_{P}(f)\}\Omega _{P}\right\Vert \rightarrow 0 \).
We obtain the relation \( \left\Vert \{Q_{P}(1,U)-1\}x\right\Vert \rightarrow 0 \)
for all \( x\in \mathcal{H}_{P} \) if \( d_{P}(U,1)\rightarrow 0 \). Thus
we have
\[
\left\Vert \{Q_{P}(1,U_{\nu })-Q_{P}(1,U)\}x\right\Vert \rightarrow 0\]
for all \( x\in \mathcal{H}_{P} \) if \( d_{P}(U_{\nu },U)\rightarrow 0 \). 

(3) Let \( f_{P}:Mp(\infty ,P)\rightarrow Sp(\infty ,P) \) be a group homomorphism
defined by
\[
f_{P}(Q_P(\lambda ,U))=U,\quad U\in Sp(\infty ,P),\quad \lambda \in \{\pm 1\}.\]
Then \( f_{P} \) is a covering map and \( \ker (f_{P})=\{Q_{P}(-1,1),Q_{P}(1,1)\} \).
Thus
\[
Mp(\infty ,P)/\ker (f_{P})\simeq Sp(\infty ,P),\]
that is, \( Mp(\infty ,P) \) is a double covering of \( Sp(\infty ,P) \). \end{proof} 

\begin{lm}\label{thm:strong_convergence_of_metaplectic_rep} \emph{Let \( Mp(\infty ,P)_{\mathrm{fin}} \)
be the group generated by \( Q_{P}(\lambda ,e^{iH}), H\in sp(\infty )\). Then the closure of \( Mp(\infty ,P)_{\mathrm{fin}} \)
with respect to the strong operator topology is \( Mp(\infty ,P) \). That is},
\emph{for any} \( U\in Sp(\infty ,P) \)\emph{,} \emph{there exists a net} \( \{U_{\mu }\} \)
\emph{in} \( Sp(\infty ) \) \emph{such that}
\[
\mathrm{s}\frac{\, \, }{\, \, }\lim _{\mu \rightarrow \infty }Q_{P}(\lambda ,U_{\mu })=Q_{P}(\lambda ,U).\]
 \end{lm}

\begin{proof} \( T(P,P') \) is written as
\[
T(P,P')=\mathrm{s}\frac{\, \, }{\, \, }\lim _{n\rightarrow \infty }Q_{P}(H(P/P')F_{n})=\mathrm{s}\frac{\, \, }{\, \, }\lim _{n\rightarrow \infty }Q_{P}(1,U(P/P')F_{n})\]
 where \( F_{n} \) is the spectral projection of a positive Hilbert-Schmidt
class operator \( \theta (P,P') \) for the open interval \( (\frac{1}{n},\infty ) \).
(See the proof of Lemma 5.5 of \cite{Araki-CCR-2}.) 

On the other hand, since \( R(U) \) is an unitary operator on \( K \), there
exists a hermitian operator \( H \) on \( K \) such that \( R(U)=e^{iH} \).
Since the set of all finite rank operators on \( K \) is dense set of all bounded
operators with respect to the strong {*} operator topology, there exists a net
\( \{A'_{\nu }\} \) such that \( A'_{\nu } \) is a finite rank operator and
\( A'_{\nu }\rightarrow H \) (strong {*} operator topology) as \( \nu \rightarrow \infty  \).
Let \( A_{\nu }:=\frac{1}{2}(A'_{\nu }+(A'_{\nu })^{*}) \). Then \( A_{\nu } \)
is a finite rank hermitian operator and \( A_{\nu }\rightarrow H \) (strong
operator topology) as \( \nu \to \infty \). Let \( H'_{\nu }:=PA_{\nu }P+(1-P)A_{\nu }(1-P) \). Then
\( H'_{\nu } \) is a finite rank hermitian operator and commutes with \( P \).
Moreover, \( H'_{\nu }\rightarrow H \) (strong operator topology) as
\( \nu \to \infty \). In fact,
for any \( x\in K \),
\[\begin{split}
\| (&H'_{\nu}-H)x \| _P\\
&\le \| (PA_{\nu}P-PHP)x\| _P+
\| ((1-P)A_{\nu}(1-P)-(1-P)H(1-P))x\| _P \\
&\le \| (A_{\nu}-H)Px\| _P+ \| (A_{\nu}-H)(1-P)x\| _P\\
&\to 0 (\nu \to \infty ).
\end{split}\]
Let \( H_{\nu }:=\frac{1}{2}(H'_{\nu }-\Gamma H'_{\nu }\Gamma ) \). Then \( H_{\nu } \)
is contained in \( sp(\infty ) \) and satisfies \( [P,H_{\nu }]=0 \) and \( H_{\nu }\rightarrow H \)
(strong operator topology) as \( \nu \rightarrow \infty  \). This shows \( e^{iH_{\nu }}\in Sp(\infty ) \)
and \( \mathrm{s}\frac{\, \, }{\, \, }\lim _{\nu \rightarrow \infty }e^{iH_{\nu }}=R(U) \).
Moreover,
\[\begin{split}
\mathrm{s}\frac{\, \, }{\, \, }\lim _{\nu \rightarrow \infty
}T_{P}(e^{iH_{\nu }})W_{P}(f)\Omega _{P} 
 & =\mathrm{s}\frac{\, \, }{\, \, }\lim _{\nu \rightarrow \infty }W_{P}(e^{iH_{\nu }}f)\Omega _{P}\\
 & = W_{P}(R(U)f)\Omega _{P}\\
 & = T_{P}(R(U))W_{P}(f)\Omega _{P}
\end{split}\]
for all \( f\in \mathrm{Re}K \). Thus
\[
\mathrm{s}\frac{\, \, }{\, \, }\lim _{\nu \rightarrow \infty }\mathrm{Q}_{P}(1,e^{iH_{\nu }})=\mathrm{s}\frac{\, \, }{\, \, }\lim _{\nu \rightarrow \infty }T_{P}(e^{iH_{\nu }})=T_{P}(R(U))=Q_{P}(1,R(U)).\]

Now let \( U_{\mu }:=U(P/P')^{\dagger }F_{n}e^{iH_{\nu }} \) where \( \mu =(\nu ,n) \).
Then
\[
Q_{P}(1,U_{\mu })=Q_{P}(1,U(P/P')^{\dagger }F_{n})Q_{P}(1,e^{iH_{\nu }})\]
and
\[
\mathrm{s}\frac{\, \, }{\, \, }\lim _{\mu \rightarrow \infty }Q_{P}(1,U_{\mu })=Q_{P}(1,U).\]
 \end{proof}

Let \( Q_{P}^{\sigma }(\lambda ,U) \) be the restriction of \( Q_{P}(\lambda ,U) \)
to \( \mathcal{H}_{P}^{\sigma } \) where \( \sigma =+ \) or \( - \).

We obtain the following proposition immediately from Lemma \ref{thm:S-S':H.S-->q(S)_q(S'):unitary_eqiv}
and Lemma \ref{thm:strong_convergence_of_metaplectic_rep}.

\begin{prop} \emph{Suppose that \( K \) is separable. Let} \( P_{1} \) \emph{and}
\( P_{2} \) \emph{be basis projections satisfying} \( \alpha \left\Vert f\right\Vert _{P_{1}}\leq \left\Vert f\right\Vert _{P_{2}}\leq \beta \left\Vert f\right\Vert _{P_{1}} \)
\emph{for all} \( f\in K \), \( K=K_{P_{1}}=K_{P_{2}} \) \emph{and} \( P_{1}-P_{2} \)
\emph{is a Hilbert-Schmidt class operator}.
\emph{Then the metaplectic representations} \( Q^{\sigma}_{P_{1}}(\lambda ,*) \) \emph{and}
\( Q^{\sigma}_{P_{2}}(\lambda ,*) \) ( \emph{resp . \(
Q^{\sigma}_{P_{1}}(\lambda ,*) \) and  \( Q^{\sigma '}_{P_{2}}(\lambda ,*) (\sigma \neq \sigma '))\)} \emph{of} \( Sp(\infty
,P_{1})=Sp(\infty ,P_{2}) \) 
\emph{are unitary equivalent}. (\emph{resp. not unitary equivalent}.) 
\end{prop}


\begin{thebibliography}{99}
\bibitem{Araki-CCR-1}Araki, Huzihiro; Shiraishi, Masafumi. 
{\em On quasifree states of the canonical commutationrelations. I. }
Publ. Res. Inst. Math. Sci. 7 (1971/72),
105--120.

\bibitem{Araki-CCR-2}Araki, Huzihiro. {\em On quasifree states of the canonical commutation relations.}
II. Publ. Res. Inst. Math. Sci. 7 (1971/72), 121--152.

\bibitem{Araki-Fock-rep-CAR}Araki, Huzihiro. 
{\em Bogoliubov automorphisms and Fock representations of canonical
anticommutation relations,} in Operator algebras and mathematical physics, 23--141,
Contemp. Math., 62, Amer. Math. Soc., Providence, RI, 1987.

\bibitem{Radon-Nikodym-theorem}Araki, Huzihiro. 
{\em Some properties of modular conjugation operator of von Neumann
algebras and a non-commutative Radon-Nikodym theorem with a chain rule.}
 Pacific J. Math. 50 (1974), 309--354.

\bibitem{Borodin-Olshanski}Borodin, Alexei; Olshanski, Grigori. 
{\em Infinite random matrices and ergodic measures.}
Comm. Math. Phys. 223 (2001), no. 1, 87--123.

\bibitem{Carey} 
Carey, A. L.; Ruijsenaars, S. N. M. 
{\em On fermion gauge groups, current algebras and Kac-Moody algebras.}
 Acta Appl. Math. 10 (1987), 
no. 1, 1--86. 

\bibitem{Lion-Vergne}Lion, G\'{e}rard; Vergne, Mic\'{e}he. 
{\em The Weil representation, Maslov index and theta series.}
 Progress in Mathematics, 6. Birkh{\"a}user, 1980.

\bibitem{Matsui1} Matsui, Taku.
{\em On quasi-equivalence of quasifree states of the gauge invariant CAR
algebras.} J.Operator Theory.17,281-290(1987)

\bibitem{Matsui}Matsui, Taku. {\em Factoriality and quasi-equivalence of quasifree states for \( Z_{2} \) and \( U(1) \) invariant CAR algebras.} 
Rev. Roumaine Math. Pures Appl. 32 (1987), no. 8, 693--700.

\bibitem{Reed-Simon-2}Michael Reed, Barry Simon. 
{\em Methods of modern mathematical physics. II. Fourier analysis, 
self-adjointness. } Academic Press, 1975.

\bibitem{Olshanski_Vershik}Olshanski, Grigori; Vershik, Anatoli. 
{\em Ergodic unitarily invariant measures on
the space of infinite Hermitian matrices.}
 Contemporary mathematical physics,
137--175, Amer. Math. Soc. Transl. Ser. 2, 175, Amer. Math. Soc., Providence,RI, 1996.

\bibitem{K.R.Parthasarathy}Parthasarathy, K. R. 
{\em An introduction to quantum stochastic calculus.}
 Birkh{\"a}user
Verlag, 1992.

\bibitem{Segal}
Pressley, Andrew; Segal, Graeme. 
{\em Loop groups.} 
Oxford Mathematical Monographs. Oxford Science Publications. 
The Clarendon Press, Oxford University Press, New York, 1986.

\bibitem{Pickrell}Pickrell, Doug. 
{\em Separable representations for automorphism groups of infinite
symmetric spaces. } J. Funct. Anal. 90 (1990), no. 1, 1--26.

\bibitem{Powers}Powers, Robert T. {\em Self-adjoint algebras of unbounded operators.} Comm. Math.Phys. 21 1971 85--124.

\bibitem{Stratila-Voiculescu}Str\u{a}til\u{a}, \c{S}erban; 
Voiculescu,Dan. {\em On a class of KMS states for
the unitary group \( \mathrm{U}(\infty ) \).} Math. Ann. 235 (1978), no. 1,
87--110.

\end{thebibliography}
\end{document}